\documentclass[
  journal=pasa,
  manuscript=research-paper, 
  year=2021,
  volume=38,
]{cup-journal}

\usepackage{microtype,siunitx,booktabs}
\usepackage{mathtools}
\usepackage{graphicx}
\usepackage{dcolumn}
\usepackage{bm}
\usepackage{amssymb}   
\usepackage{hyperref}
\usepackage[caption=false]{subfig}
\usepackage{xcolor}

\newcommand\farcs{\hbox{$.\!\!^{\prime\prime}$}}

\newcommand\arcdeg{\mbox{$^\circ$}}

\title{A Geometric View of Closure Phases in Interferometry}

\author{Nithyanandan Thyagarajan}
\affiliation{National Radio Astronomy Observatory, 1003 Lopezville Road, Socorro, NM 87801, USA}
\alsoaffiliation{Commonwealth Scientific and Industrial Research Organisation (CSIRO), Space \& Astronomy, P. O. Box 1130, Bentley, WA 6102, Australia}
\email[Nithyanandan Thyagarajan]{Nithyanandan.Thyagarajan@csiro.au}

\author{Chris L. Carilli}
\affiliation{National Radio Astronomy Observatory, 1003 Lopezville Road, Socorro, NM 87801, USA}

\received{\today}
\revised{\today}
\accepted{\today}
\published{\today}


\abbreviations{
  AGN: Active Galactic Nucleus,
  ALMA: Atacama Large Millimeter Array,
  EHT: Event Horizon Telescope,
  EM: Electro-magnetic,
  FR II: Fanaroff-Riley II,
  LMT: Large Millimeter Telescope,
  SMA: Sub-millimeter Array,
  NPC: Null Phase Curve,
  SOS: Shape-orientation-size,
  VLA: Karl~G.~Jansky Very Large Array,
  VLBI: Very Long Baseline Interferometry
}

\begin{document}

\begin{abstract}

Closure phase is the phase of a closed-loop product of spatial coherences formed by a $\ge 3$-element interferometer array. Its invariance to phase corruption attributable to individual array elements acquired during the propagation and the measurement processes, subsequent calibration, and errors therein, makes it a valuable tool in interferometry applications that otherwise require high-accuracy phase calibration. However, its understanding has remained mainly mathematical and limited to the aperture plane (Fourier dual of the image plane). Here, we present a geometrical, image-domain view of closure phase, which until now has been lacking. Using the principal triangle in a 3-element interference image formed by a triad of interferometer elements, we show that the properties of closure phase, particularly its invariance to multiplicative element-based corruption factors (even of a large magnitude) and to translation, are intricately related to the conserved properties of the triangle, namely, its shape, orientation, and size, which is referred herein as the ``shape-orientation-size (SOS) conservation principle''. In the absence of a need for element-based amplitude calibration of the interferometer array (as is typical in optical interferometry), the principal triangle in any 3-element interference image formed from phase-uncalibrated spatial coherences is still a true and uncorrupted representation of the source object's morphology, except for a possible shift. Based on this knowledge of the triangle SOS conservation principle, we present two geometric methods to measure the closure phase directly from a simple 3-element interference image (without requiring an aperture-plane view): (i) the closure phase is directly measurable from any one of the triangle's heights, and (ii) the squared closure phase is proportional to the product of the areas enclosed by the triad of array elements and the principal triangle in the aperture and image planes, respectively. 
We validate the geometric understanding of closure phase in the image plane using observations with the Karl~G.~Jansky Very Large Array, and the Event Horizon Telescope. These results verify the SOS conservation principle across a wide range of radio interferometric conditions. This geometric insight can be potentially valuable to other interferometric applications, such as optical interferometry. We also generalise these geometric relationships to an $N$-element interferometer. 

\end{abstract}


\section{INTRODUCTION} \label{sec:intro}

The concept of closure phase in radio interferometry can be traced back to \citet{jen58}. Closure phase provides information on the phase encoded in the spatial coherences due only to the intensity distribution of sources of electromagnetic (EM) radiation in the sky, without the need for calibration to correct for corruption of the phases of the EM waves due to propagation effects and the array receiver elements themselves. The invariance of the closure phase to phase corruptions of the incident EM wave that can be factorised into element-based phase terms, has been extensively tested and applied in interferometry. This property has played a significant role in the development of a popular calibration scheme called ``self-calibration'' \citep{Schwab1980,Cornwell+1981,Pearson+1984}. Moreover, closure phase is known for its measure of the \textit{centrosymmetry} or \textit{point-symmetry} (morphological symmetry around a point) as well as for its invariance to translation of the spatial intensity distribution of the EM radiation \citep{mon07b}. A general theory of identifying a complete and independent set of closure invariants from co-polar and polarimetric interferometry is beginning to emerge \citep{Broderick+2020,Thyagarajan+2021a,Samuel+2021}. 

The invariance of closure invariants to local, element-based corruptions have made them a valuable tool in experiments that face challenges due to the requirement of high-accuracy calibration. Closure phases have thus been used in optical interferometry to characterise the structures of stars \citep[][and references therein]{mon03a,mon03b,mon06,mon07a}, discoveries with Very Long Baseline Interferometry (VLBI) such as the core-jet morphology of the quasar 3C~147 \citep{Wilkinson+1977}, the detection of the superluminal expansion of the relativistic jet in quasar 3C~273 \citep{Pearson+1981}, and the Event Horizon Telescope (EHT) imaging of the shadow of the supermassive black hole in M87 \citep{eht19-1,eht19-2,eht19-3,eht19-4,eht19-5,eht19-6,eht21-7} and Centaurus~A \citep{Janssen+2021}. Recently, closure phase has provided a useful avenue towards detecting the neutral Hydrogen structures during the \textit{cosmic reionisation} (at redshifts, $z\gtrsim 6$) using its characteristic 21~cm spectral line redshifted to low radio frequencies with interferometer arrays \citep{car18,car20,thy18,thy20a,thy20b}. 

However, despite extensive use and successful applications spanning several decades, a geometric insight into the interferometric closure phase has remained elusive. The complex, higher-order dependence on the moments of the spatial intensity and spatial coherence \citep{lac03,thy20a} makes it very challenging to gain a geometric intuition of this special quantity. In this paper, we address this issue by developing a method to visualise the manifestation of closure phase in 3-element interference images, and present two methods to estimate the closure phase therein. While closure phase can always be calculated from three separate measurements of the offsets (or equivalently, phases) of individual fringes from a reference point in an image, our methods generally require only a single measurement on a three fringe image except in special cases. We anticipate that this insight will result in widening the spectrum of synthesis interferometry applications.

The paper is organised as follows. \S\ref{sec:interferometry} sets up the interferometry context and introduces the closure phase of an $N$-polygon interferometer array. In \S\ref{sec:geometric-view}, we present the geometrical characteristics, and direct geometrical methods for the estimation of closure phase in the image plane using a 3-element interferometer, through a derivation of the shape-orientation-size conserving property of closure phase. A validation via applications to real radio interferometric data from observations of bright cosmic objects at centimeter wavelengths using the Karl~G.~Jansky Very Large Array (VLA) radio telescope, as well as in the VLBI regime at millimeter wavelengths using publicly available EHT data is provided in \S\ref{sec:data}. The findings are summarised in \S\ref{sec:summary}. 

\section{THE INTERFEROMETRY CONTEXT} \label{sec:interferometry}

\begin{figure}
\includegraphics[width=\linewidth]{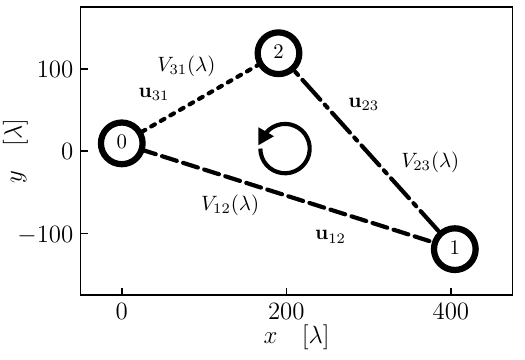}
\caption{A triad of aperture elements with positions, $\boldsymbol{x}_a/\lambda$, and spacings, $\boldsymbol{u}_{ab}$, both in units of wavelengths, with $a,b=0, 1, 2$, and $b\ne a$. $\boldsymbol{u}_{ab}$ represents the spatial frequencies of the image-plane intensity distribution, $I(\hat{\boldsymbol{s}},\lambda)$, in the aperture plane. $V_{ab}(\lambda)$ denotes the complex-valued spatial coherence of $I(\hat{\boldsymbol{s}},\lambda)$ measured at $\boldsymbol{u}_{ab}$ in the aperture plane. The cyclic ordering of the element spacings is indicated by the arrowed (anti-clockwise) circle. The three spatial frequencies, $\boldsymbol{u}_{ab}$, are shown by dashed, dash-dotted, and dotted lines, which will be used to denote the corresponding  fringes in the image plane in subsequent figures.
\label{fig:antpos}}
\end{figure}

Consider measurements of a single polarisation state of a complex-valued, quasi-monochromatic electric field, $E_a(\lambda)$, integrated over a narrow band around the wavelength, $\lambda$, of the incident EM radiation by $N_\textrm{A} $ array elements at locations $\boldsymbol{x}_a$, with $a=0, 1, \ldots N_\textrm{A}-1$ in the aperture plane. The spacing between any pair of array elements (commonly referred as to as the baseline vector in radio interferometry) is denoted by $\boldsymbol{x}_{ab} \equiv \boldsymbol{x}_b - \boldsymbol{x}_a$. The spatial distribution of the intensity of the EM radiation in the image plane, $I(\hat{\boldsymbol{s}},\lambda)$, and the corresponding spatial coherence of the electric fields (also known as \textit{visibilities} in radio interferometry) in the aperture plane, $V_{ab}(\lambda)$, exhibit a Fourier-transform relationship with each other \citep{Born-Wolf-1999,TMS2017,SIRA-II},
\begin{align}
    V_{ab}(\lambda) &\coloneqq \left\langle E_a^\star(\lambda)E_b(\lambda)\right\rangle \label{eqn:E-field-corr} \\
    &= \int_\Omega \Theta(\hat{\boldsymbol{s}},\lambda)\, I(\hat{\boldsymbol{s}},\lambda)\, e^{-i 2\pi \boldsymbol{u}_{ab}\cdot \hat{\boldsymbol{s}}}\, \mathrm{d}\Omega \, , \label{eqn:vis-img-FT}
\end{align}
where, the angular brackets, $\langle\cdot\rangle$, represent a true ensemble average, $\hat{\boldsymbol{s}}$ denotes a unit vector in the direction of any location in the image, $\boldsymbol{u}_{ab}\coloneqq\boldsymbol{x}_{ab}/\lambda\equiv (u_{ab}, v_{ab}, w_{ab})$ defined on the aperture plane denotes the array element spacings with $u_{ab}$ and $v_{ab}$ being projections on the plane perpendicular to, and $w_{ab}$ being towards the direction of the phase centre, $\hat{\boldsymbol{s}}_0$, in the image. In the Fourier relationship, $\boldsymbol{u}_{ab}$, by definition, represents the spatial frequencies of the structures in $I(\hat{\boldsymbol{s}},\lambda)$. The array element's directional power pattern is denoted by $\Theta(\hat{\boldsymbol{s}},\lambda)$, and $\mathrm{d}\Omega$ denotes the differential solid angle in the image plane perpendicular to $\hat{\boldsymbol{s}}$. The vectors $\hat{\boldsymbol{s}}$ and $\boldsymbol{u}_{ab}$ can be represented on a Cartesian coordinate frame with orthogonal basis vectors, $\hat{\boldsymbol{e}}_x$, $\hat{\boldsymbol{e}}_y$, and $\hat{\boldsymbol{e}}_z$. In this frame, $\hat{\boldsymbol{s}}\equiv \ell\,\hat{\boldsymbol{e}}_x + m\,\hat{\boldsymbol{e}}_y + n\hat{\boldsymbol{e}}_z$ with $\ell^2 + m^2 + n^2 = 1$, where $l$, $m$, and $n$ denote the direction-cosines of $\hat{\boldsymbol{s}}$. And, $\boldsymbol{u}_{ab}\equiv u_{ab}\,\hat{\boldsymbol{e}}_x + v_{ab}\,\hat{\boldsymbol{e}}_y + w_{ab}\hat{\boldsymbol{e}}_z$. 
Figure~\ref{fig:antpos} depicts the modeled locations of three array elements in units of wavelengths (chosen at $\lambda=21$~cm) that will be used in the initial examples that follow. The cyclic ordering of the element indices is indicated by the arrowed circle. The three encircled elements can be considered as three antennas in a radio interferometer, or optical mirrors or aperture mask openings in an optical interferometer. 

In practice, the EM voltage measurements at the array elements are inevitably corrupted by multiplicative complex-valued ``gain'' factors introduced by the intervening medium as well as the array element response. The corrupted measurements, either pre-calibration or after miscalibration, are denoted by $\widetilde{E}_a(\lambda) = G_a(\lambda)\, E_a(\lambda)$, where, $G_a(\lambda)$ denotes the net corruption factors introduced in the measurement process factorisable in such a way that it is attributable to the individual elements. The process of calibration \citep{TMS2017,SIRA-II} aims to undo this corruption. In this paper, $G_a(\lambda)$ refers to the net corruption in a measurement, whether it is uncalibrated or imperfectly calibrated.

The corrupted (uncalibrated or miscalibrated) visibility is 
\begin{align}
    \widetilde{V}_{ab}(\lambda) &= G_a^\star(\lambda) G_b(\lambda) V_{ab}(\lambda) \label{eqn:E-field-cal-corr} \\ 
    &= \left|G_a^\star(\lambda) G_b(\lambda) V_{ab}(\lambda) \right| \, e^{i[\xi_b(\lambda)-\xi_a(\lambda) + \phi_{ab}(\lambda)]}
 \label{eqn:calvis-img-FT}
\end{align}
where, $\xi_a(\lambda)=\arg G_a(\lambda)$, $\phi_{ab}(\lambda)=\arg V_{ab}(\lambda)$, and 
\begin{align}\label{eqn:radio-interferometry-phase-corruption}
    \widetilde{\phi}_{ab}(\lambda)=\arg \widetilde{V}_{ab}(\lambda) &= \phi_{ab}(\lambda) + \xi_b(\lambda)-\xi_a(\lambda) \, .
\end{align}
The residual gain in an ideal measurement without corruption or after a perfect calibration is, $G_a(\lambda)=1$, and thus $\xi_a(\lambda)=0$ and $\widetilde{\phi}_{ab}(\lambda)=\phi_{ab}(\lambda)$ for all $a$ and $b$. However, it is often difficult to realise in practice. 

In the rest of the paper, we assume a flat image plane without significant curvature effects, usually referred to as narrow field of view or ``flat sky'' approximation ($\ell,m\ll 1$) in radio interferometry. In such a scenario, $w_{ab}\ne 0$ will only result in a translation of the images as per the extra phase term in the visibility arising from the additional path length due to $w_{ab}$ in Equation~(\ref{eqn:vis-img-FT}), and will not affect their geometry or morphology. Therefore, we choose an aperture plane which is coplanar ($w_{ab}=0$). The effects of non-coplanarity in the presence of significant image plane curvature will be the subject of future study. 

\subsection{Interferometric Fringes}\label{sec:fringes}

The image plane response of a single interferometer (visibility measured on one baseline) corresponding to the correlator output or spatial coherence, is called a ``fringe'',which for the general corrupted form is
\begin{align}\label{eqn:fringe}
    \widetilde{F}_{ab}(\hat{\boldsymbol{s}},\lambda) &= \left|\widetilde{V}_{ab}(\lambda)\right| \, e^{i [2\pi \boldsymbol{u}_{ab}\cdot \hat{\boldsymbol{s}} + \phi_{ab}(\lambda) + \xi_b(\lambda)-\xi_a(\lambda)]} 
\end{align}
with $\arg \widetilde{F}_{ab}(\hat{\boldsymbol{s}},\lambda) = 2\pi \boldsymbol{u}_{ab}\cdot \hat{\boldsymbol{s}} + \phi_{ab}(\lambda) + \xi_b(\lambda)-\xi_a(\lambda)\,$. Ideally, $\arg \widetilde{F}_{ab}(\hat{\boldsymbol{s}},\lambda)=\arg F_{ab}(\hat{\boldsymbol{s}},\lambda) = 2\pi \boldsymbol{u}_{ab}\cdot \hat{\boldsymbol{s}} + \phi_{ab}(\lambda)$ occurs when all elements are perfectly phase-calibrated, that is, $\xi_a(\lambda)=0$ for all $a$.

These fringes are periodic, marked by ``ridges'' corresponding to where $\arg \widetilde{F}_{ab}(\hat{\boldsymbol{s}},\lambda)$ is an integer multiple of $2\pi$, and are given by the null-valued (or zero-valued) isophase contours\footnote{Any (but same) constant-valued phase contours can be used for all fringes, but without loss of generality, we use zero-valued contours corresponding to the ridges in the fringes, for convenience.} (and their equivalents offset by multiples of $2\pi$). These are, hereafter, called the fringe null phase curves (NPC):
\begin{align}\label{eqn:npc}
    2\pi \boldsymbol{u}_{ab}\cdot \hat{\boldsymbol{s}} + \widetilde{\psi}_{ab}(\lambda) &= 0 \, , \quad a,b=0,\ldots N-1 \, , \, \textrm{with}\, a\ne b \, ,
\end{align}
where, $\widetilde{\psi}_{ab}(\lambda) = \widetilde{\phi}_{ab}(\lambda) + 2\pi n_{ab}\,$, and $n_{ab}$ (an integer) accounts for the NPC offset from the principal NPC ($n_{ab}=0$) by integer multiples ($n_{ab}$) of $2\pi$. When traversing anywhere on an NPC (ridge), $\arg \widetilde{F}_{ab}(\hat{\boldsymbol{s}},\lambda)$ remains unchanged.

The signed positional offset, $\delta s_{ab}(\lambda)$, of the fringe NPC from the phase centre (origin) along a perpendicular and the corresponding phase offset, $\widetilde{\psi}_{ab}(\lambda)$, are related by
\begin{align}\label{eqn:perpendicular-phase-offset}
    \widetilde{\psi}_{ab}(\lambda) &= 2\pi \left|\boldsymbol{u}_{ab}\right|\,\delta s_{ab}(\lambda) \, .
\end{align}
Here, $|\boldsymbol{u}_{ab}|=(u_{ab}^2+v_{ab}^2)^{1/2}$ denotes the baseline lengths in the projected plane whose normal is the direction to the phase centre. 
Because $\boldsymbol{u}_{ab}$ is the spatial frequency of a fringe, $1/|\boldsymbol{u}_{ab}|$ represents the spatial period of the periodic ridges (or the fringe spacing) in the image plane and corresponds to a phase change of $2\pi$, as verified by setting $\delta s_{ab}(\lambda)=1/|\boldsymbol{u}_{ab}|$ in Equation~(\ref{eqn:perpendicular-phase-offset}).

\begin{figure*}
\includegraphics[width=\linewidth]{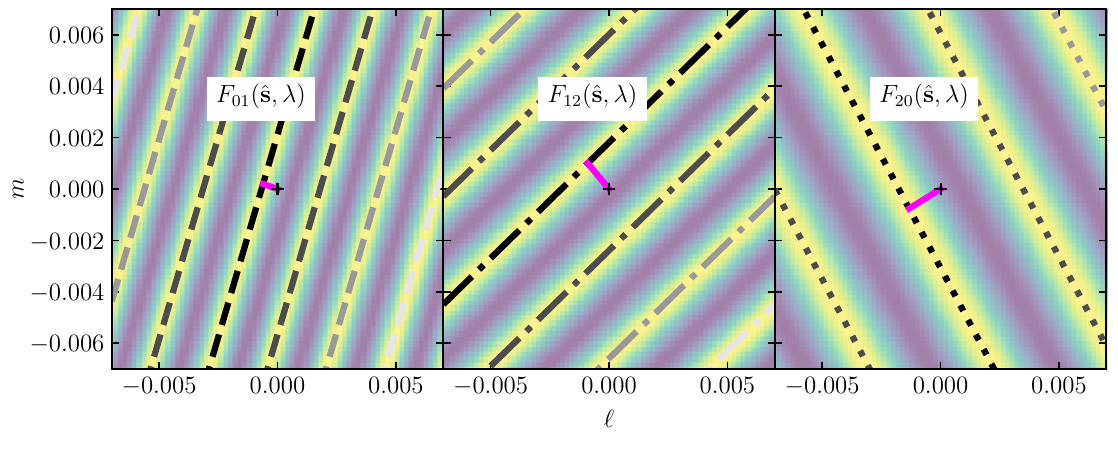}
\caption{Ideal fringes, $F_{ab}(\hat{\boldsymbol{s}},\lambda)$, and the respective NPCs (lines) in the image plane in direction-cosine ($\ell, m$) coordinates, with the line style in each panel corresponding to that of the array element spacings, $\boldsymbol{u}_{ab}$, in Figure~\ref{fig:antpos}. Equation~(\ref{eqn:npc}) yields the fringe NPCs. The black lines in each line style corresponds to the principal fringe NPC ($n_{ab}=0$), while the varying shades of gray correspond to secondary ($|n_{ab}|>0$) fringe NPCs. The phase centre (origin) is marked (with a $+$ symbol). The positional offset from the phase centre to each of the principal fringe NPCs is shown in magenta and is related to the visibility phase, $\phi_{ab}(\lambda)$, by Equation~(\ref{eqn:perpendicular-phase-offset}). \label{fig:triad-fringes-v2}}
\end{figure*}

Figure~\ref{fig:triad-fringes-v2} shows the ideal fringes, $F_{ab}(\hat{\boldsymbol{s}},\lambda)$, in the image plane in direction-cosine coordinates, ($\ell, m$), given by Equation~(\ref{eqn:fringe}) for the modeled 3-element array and the corresponding visibilities shown in Figure~\ref{fig:antpos}. The $+$ symbol marks the phase centre (origin). The fringe NPCs, described by Equation~(\ref{eqn:npc}), are shown in line styles corresponding to those in Figure~\ref{fig:antpos}. The black line in each panel denotes the principal NPC ($n_{ab}=0$) of the corresponding fringe. The various gray lines denote the secondary NPCs ($|n_{ab}|>0$) of the fringes. The positional offset, $\delta s_{ab}(\lambda)$, of the principal fringe NPC from the phase centre is shown by the magenta segments and corresponds to $\phi_{ab}(\lambda)$ (the principal visibility phase) according to Equation~(\ref{eqn:perpendicular-phase-offset}). In the case of uncalibrated or imperfectly calibrated visibilities, these phase offsets also include the corruptions, $\xi_b(\lambda)-\xi_a(\lambda)$, arising from the complex gains of the array elements. 

\subsection{Interferometric Closure Phase}\label{sec:cphase}

Hereafter, we will assume that the angular power patterns, $\Theta(\hat{\boldsymbol{s}},\lambda)$, of the array elements are identical. Consider $N$ elements forming an $N$-vertex polygon in the aperture plane. The element spacings in the adjacent sides in the polygon are given by $\boldsymbol{x}_{a\lceil a+1\rfloor_N} \equiv \boldsymbol{x}_{\lceil a+1\rfloor_N} - \boldsymbol{x}_a$, where, $\lceil a\rfloor_N \equiv a \mod N$. The condition for a closed loop in the aperture plane is
\begin{align}\label{eqn:closed-polygon}
    \sum_{a=0}^{N-1} \boldsymbol{u}_{a\lceil a+1\rfloor_N} &\equiv \boldsymbol{0} \, .
\end{align}

The interferometric closure phase on the $N$-polygon is
\begin{align}
    \widetilde{\phi}_N(\lambda) &\equiv \arg\prod_{a=0}^{N-1} \widetilde{V}_{a\lceil a+1\rfloor_N}(\lambda) = \sum_{a=0}^{N-1} \arg\widetilde{V}_{a\lceil a+1\rfloor_N}(\lambda) \, .
\end{align}
Because $\sum_{a=0}^{N-1} \arg G_a^\star(\lambda) \, G_{\lceil a+1\rfloor_N}(\lambda) \equiv 0\,$, 
\begin{align}\label{eqn:CPhase-N-polygon}
    \widetilde{\phi}_N(\lambda) &= \phi_N(\lambda) = \sum_{a=0}^{N-1} \arg V_{a\lceil a+1\rfloor_N}(\lambda) \, ,
\end{align}
where, $\phi_N(\lambda)$ is the true closure phase on the $N$-polygon. Therefore, the closure phase is invariant to element-based corruptions, $G_a(\lambda)$, the corrections from calibration, as well as the errors therein, making it a true observable physical property of the structures in the image-plane intensity distribution. This property is a form of \textit{gauge-invariance} with respect to any element-based phases acquired during the measurement process \citep{Thyagarajan+2021a,Samuel+2021}. 

A direct consequence of this gauge-invariance is that the closure phase is also invariant to translation in the image plane. This can be shown by replacing $\hat{\boldsymbol{s}}$ with $\hat{\boldsymbol{s}}^\prime = \hat{\boldsymbol{s}} - \hat{\boldsymbol{s}}_0$, where $\hat{\boldsymbol{s}}_0$ is an arbitrary choice for the origin of the image plane, referred to as the \textit{phase centre} in interferometry. From Equation~(\ref{eqn:vis-img-FT}), such a translation modifies the spatial coherence as
\begin{align}
    V_{ab}^\prime(\lambda) &= e^{i 2\pi \boldsymbol{u}_{ab}\cdot \hat{\boldsymbol{s}}_0}\, V_{ab}(\lambda) \, , \label{eqn:translation-phase}
\end{align}
which simply introduces an additional phase factor, $e^{i 2\pi \boldsymbol{u}_{ab}\cdot \hat{\boldsymbol{s}}_0}$,
that is factorisable into element-based phase factors, $e^{i 2\pi\boldsymbol{x}_b\cdot \hat{\boldsymbol{s}}_0 /\lambda}$ and $e^{-i 2\pi\boldsymbol{x}_a\cdot \hat{\boldsymbol{s}}_0 /\lambda}$. Due to the gauge-invariance discussed above, the closure phase is therefore independent of the phase factors introduced by translation in the image plane. Conversely, the translation invariance of the closure phase is simply a special case of the gauge-invariance to the phase factors attributable locally to the array elements.

Because closure phases are unaffected by the corrupting gain terms as is well-known and as shown above, without loss of generality, we choose to work with uncorrupted terms in subsequent mathematical analysis unless specified otherwise.

\section{SHAPE-ORIENTATION-SIZE (SOS) CONSERVATION}\label{sec:geometric-view}

In the following, we mathematically derive, and then demonstrate with model and real data, the underlying geometric nature of closure phase using the image-plane fringes of a closed triad of array elements, in the case when visibility phase corruption can be assigned to individual elements, and is not idiosyncratic to a given baseline. The equivalent geometric behavior manifests as a `shape-orientation-size' (SOS) conservation, in which the shape, orientation, and size of the triangle enclosed by the three principal fringe NPCs from a closed triad of array elements are conserved even in the presence of large phase errors, except possibly an overall translation in the image plane that does not affect the SOS conservation.

A triad ($N=3$), being the simplest closed shape for studying the closure phase, will form the basis later for characterising the behavior on $N$-polygons with $N>3$. Consider three fringes $F_{a\lceil a+1\rfloor_N}(\hat{\boldsymbol{s}},\lambda)$ with $N=3$ and $a=0,1,2$. The corresponding NPCs are obtained from Equation~(\ref{eqn:npc}) as
\begin{align}\label{eqn:triad-npc}
    2\pi \boldsymbol{u}_{a\lceil a+1\rfloor_3}\cdot \hat{\boldsymbol{s}} + \psi_{a\lceil a+1\rfloor_3}(\lambda) &= 0 \, , \quad a=0,1,2 \, .
\end{align}
They enclose a ``principal'' triangle\footnote{The principal fringe NPCs enclose the principal triangle.}, in general\footnote{Collinear array elements will yield coincident or non-intersecting fringes, which will enclose a flattened degenerate triangle.}.

In Equation~(\ref{eqn:triad-npc}), $\psi_{a\lceil a+1\rfloor_3}(\lambda)$ is the phase offset from the phase centre, which has been implicitly assumed to be at $\hat{\boldsymbol{s}}_0\equiv (0,0,1)$. Thus, the closure phase on the 3-polygon is
\begin{align}\label{eqn:CPhase-3-sum}
    \psi_3(\lambda) &\equiv \sum_{a=0}^2 \psi_{a\lceil a+1\rfloor_3}(\lambda) \, ,
\end{align}
which is simply the sum of the phase offsets of the individual fringe NPCs from the phase centre. Geometrically, the phase offsets are obtained from Equation~(\ref{eqn:perpendicular-phase-offset}) by measuring the positional offsets from the phase centre to each of these fringe NPC [Equation~(\ref{eqn:triad-npc})] normalised by the respective fringe spacings. For a calibrated interferometer, these measured phase offsets relate directly to the object's position and structure.

If the phase centre is shifted to some arbitrary $\hat{\boldsymbol{s}}_0$, then by defining $\hat{\boldsymbol{s}}^\prime = \hat{\boldsymbol{s}} - \hat{\boldsymbol{s}}_0$, Equation~(\ref{eqn:triad-npc}) for the fringe NPCs will be
\begin{align}\label{eqn:modified-npc}
    2\pi \boldsymbol{u}_{a\lceil a+1\rfloor_3}\cdot \hat{\boldsymbol{s}}^\prime + \psi_{a\lceil a+1\rfloor_3}^\prime(\lambda) &= 0 \, , \quad a=0,1,2 \, .
\end{align}
Then, the closure phase with the shifted phase centre is
\begin{align}\label{eqn:modified-CPhase-3-sum}
    \psi_3^\prime(\lambda) &\equiv \sum_{a=0}^2 \psi_{a\lceil a+1\rfloor_3}^\prime(\lambda) \nonumber\\
    &= \sum_{a=0}^2 \psi_{a\lceil a+1\rfloor_3}(\lambda) + 2\pi \hat{\boldsymbol{s}}_0\cdot \sum_{a=0}^2 \boldsymbol{u}_{a\lceil a+1\rfloor_3} \nonumber\\
    &= \psi_3(\lambda) \, ,
\end{align}
where, we have used Equations~(\ref{eqn:closed-polygon}) and (\ref{eqn:CPhase-3-sum}). This reiterates, using an image-based geometric viewpoint, the common knowledge that the closure phase remains invariant despite an arbitrary translation. 

\subsection{Relation to the Height of the Principal Triangle}\label{sec:phase-center-triads}

Consider the principal triangle enclosed by the three principal NPCs. The phase centre, $\hat{\boldsymbol{s}}_0$, can be conveniently chosen to be at any of the vertices of this triangle, which is the point of intersection of any of the two fringe NPCs, for instance, $F_{01}(\hat{\boldsymbol{s}},\lambda)$ and $F_{12}(\hat{\boldsymbol{s}},\lambda)$. Because $\hat{\boldsymbol{s}}_0$ lies on the NPCs of both $F_{01}(\hat{\boldsymbol{s}},\lambda)$ and $F_{12}(\hat{\boldsymbol{s}},\lambda)$, by definition, $\delta s_{01}^\prime(\lambda)= \delta s_{12}^\prime(\lambda) = 0\,$, and therefore, $\psi_{01}^\prime(\lambda) = \psi_{12}^\prime(\lambda) = 0$ from Equation~(\ref{eqn:perpendicular-phase-offset}). Hence,
\begin{align}\label{eqn:CPhase-opp-vertex-intersection}
    \psi_3^\prime(\lambda) &= \psi_{20}^\prime(\lambda) = \psi_{20}(\lambda) + 2\pi \boldsymbol{u}_{20}\cdot \hat{\boldsymbol{s}}_0 \nonumber\\
    &= \psi_{20}(\lambda) - 2\pi (\boldsymbol{u}_{01}+\boldsymbol{u}_{12})\cdot \hat{\boldsymbol{s}}_0 \nonumber\\
    &= \sum_{a=0}^2 \psi_{a\lceil a+1\rfloor_3}(\lambda) = \psi_3(\lambda) \, .
\end{align}
Thus, when the phase centre is chosen to be at any of the vertices of the principal triangle, the closure phase is simply
\begin{align}\label{eqn:perpendicular-phase-offset-vertex}
    \psi_3(\lambda) &= \psi_{a\lceil a+1\rfloor_3}^\prime(\lambda) = 2\pi \left|\boldsymbol{u}_{a\lceil a+1\rfloor_3}\right|\,\delta s_{a\lceil a+1\rfloor_3}^\prime(\lambda)
\end{align}
from Equations~(\ref{eqn:perpendicular-phase-offset}) and (\ref{eqn:CPhase-opp-vertex-intersection}), where, $\delta s_{a\lceil a+1\rfloor_3}^\prime(\lambda)$ is the height drawn from the vertex chosen as the phase centre, to the opposite side corresponding to the fringe NPC, $F_{a\lceil a+1\rfloor_3}(\hat{\boldsymbol{s}},\lambda)$. Thus, the closure phase is directly calculable from a single measurement of any one of the heights\footnote{When the array elements are arranged collinearly on the aperture plane, the resulting fringes are all parallel to each other yielding no definite intersections between the fringe NPCs that could serve as the preferred phase centres. However, the closure phase is still well-defined. An arbitrary phase centre can be still chosen, including anywhere on one of the fringe NPCs, and the closure phase is given by Equation~(\ref{eqn:modified-CPhase-3-sum}) as in the general case.} of this triangle.

Invariance of closure phase to element-based gain corruptions then implies the heights of the triangle also have to remain invariant. And because the element spacings (baseline vectors) do not change during the measurement process, the orientations of the sides of the triangle in the image plane are fixed. Therefore, all three characteristics, namely, shape, orientation, and size of the triangle in the 3-element interference image are preserved. We will refer to this as the shape-orientation-size (SOS) conservation principle. The only remaining degree of freedom for this triangle that does not affect its SOS characteristics is an overall translation.

\begin{figure*}
\centering
\subfloat[][Closure phase from ideal fringes \label{fig:ideal-fringes-closure-phase}]{\includegraphics[width=0.365\textwidth]{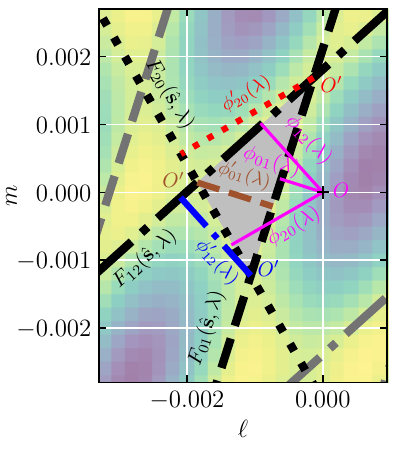}}
\subfloat[][Invariance of closure phase from uncalibrated and translated fringes\label{fig:uncal-trans-fringes-closure-phases}]{\includegraphics[width=0.63\textwidth]{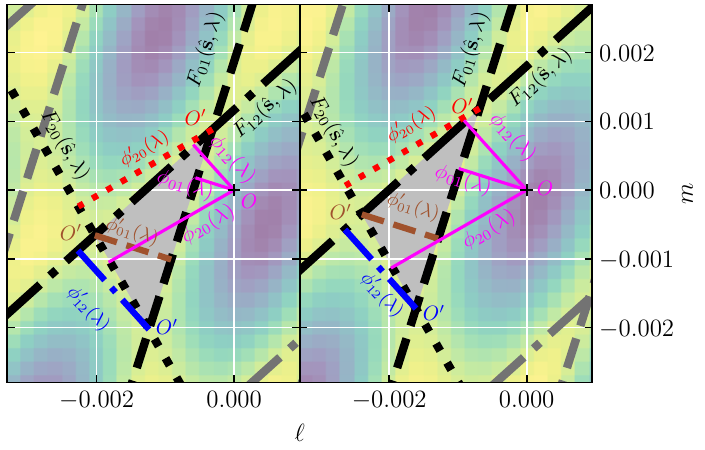}}
\caption{Illustration of the gauge-invariant and shape-orientation-size (SOS) conserving nature of closure phase. (a)~Visibility fringes and phases, and closure phase on ideal (or perfectly calibrated) fringes, $F_{ab}(\hat{\boldsymbol{s}},\lambda)$ for $a = 0, 1, 2$, $b = \lceil a+1\rfloor_3$. The three principal fringe NPCs are annotated and shown in black lines with the line style corresponding to that in Figures~\ref{fig:antpos} and \ref{fig:triad-fringes-v2}. They enclose the principal triangle marked by the gray shaded region. Gray lines denote secondary fringe NPCs. The three principal  visibility phases, $\phi_{ab}(\lambda)$, are proportional to the positional offsets [see Equation~(\ref{eqn:perpendicular-phase-offset})] shown in magenta from the phase centre (origin) marked by $+$ and annotated by $\mathcal{O}$. The closure phase from the principal fringes, $\phi_3(\lambda)$, is the sum of the three visibility phases. The phase centre can be conveniently shifted to any one of the triangle's vertices, $\mathcal{O}^\prime$, marked in brown, blue, or red, in which case the closure phase reduces simply to $\phi_{ab}^\prime(\lambda)$, which are shown corresponding to the heights drawn from the vertex to the opposite side in brown (dashed), blue (dash-dotted), or red (dotted), respectively, according to Equation~(\ref{eqn:perpendicular-phase-offset-vertex}). Moreover, the area enclosed by the triangle is proportional to the closure phase squared (see \S\ref{sec:areas-triad}). (b)~Same as the ideal case in panel (a) but when considering uncalibrated (all three element phases corrupted randomly ranging from 15\arcdeg~ to 75\arcdeg) and translated fringes in the middle and right panels, respectively. As a result, all the fringe NPCs are displaced parallel to themselves relative to the phase centre compared to the ideal case. The closure phase, which is still the sum of the three uncalibrated or translated visibility phases (corresponding to the positional offsets in magenta), remains unchanged. The geometrical equivalence is that the closure phase which is proportional to the heights drawn from one of the triangle's vertices to the corresponding opposite side (brown dashed, blue dash-dotted, or red dotted lines) are independent of these shifts as well as of the phase centre. Though the individual fringes and the triangle enclosed by them are displaced, their displacements are constrained to be parallel to themselves with the only degree of freedom being an overall translation of the triangle, thereby conserving its SOS characteristics (hence, the area too). The SOS conservation, despite electromagnetic phase corruption attributable to individual array elements, and an overall translation in the image plane, geometrically demonstrates the gauge-invariance of closure phase. \label{fig:gauge-invariance}}
\end{figure*}

Figure~\ref{fig:ideal-fringes-closure-phase} illustrates these relations geometrically using image obtained by interfering the three fringes, $F_{a\lceil a+1\rfloor_3}(\hat{\boldsymbol{s}},\lambda)$, shown in Figure~\ref{fig:triad-fringes-v2}. The black and gray lines denote the principal and secondary NPCs of the fringes, respectively, with line styles corresponding to those in Figures~\ref{fig:antpos} and \ref{fig:triad-fringes-v2}. The $+$ symbol marks the phase centre (or the origin) and is denoted by $\mathcal{O}$ in magenta. The principal triangle is shown by the gray shaded region. The positional offsets, $\delta s_{a\lceil a+1\rfloor_3}(\lambda)$, of the principal fringe NPCs from the phase centre are shown as magenta lines annotated by the corresponding principal visibility phases, $\phi_{a\lceil a+1\rfloor_3}(\lambda)$, obtained using Equation~(\ref{eqn:perpendicular-phase-offset}). When the phase centre is conveniently chosen to be at any of the vertices of the triangle (denoted by $\mathcal{O}^\prime$ in red, blue, and brown), the modified visibility phases, $\phi_{a\lceil a+1\rfloor_3}^\prime(\lambda)$, are proportional to the heights of the triangle, $\delta s_{a\lceil a+1\rfloor_3}^\prime(\lambda)$, drawn from the chosen vertex (phase centre) shown by the corresponding colored lines, according to Equation~(\ref{eqn:perpendicular-phase-offset-vertex})\footnote{We note that, for a given vertex, there can be flipped or complementary triangles in the image plane from which the closure phase can be derived. Two of these can be seen to the left and right of the brown $\mathcal{O}^\prime$ vertex in Figure~\ref{fig:ideal-fringes-closure-phase}. The sum of the two closure phases from the complementary triangles sharing a vertex must be, by definition, $2\pi$, thereby demonstrating the $2\pi$ ambiguity of phase, encapsulated by $n_{ab}$ following Equation~(\ref{eqn:npc}).}. The same equation also implies that each of these modified principal visibility phases, $\phi_{a\lceil a+1\rfloor_3}^\prime(\lambda)$, is equal to the principal closure phase, $\phi_3(\lambda)$, or in general, $\psi_3(\lambda) = \psi_{a\lceil a+1\rfloor_3}^\prime(\lambda), \forall a$ when the $2\pi$ phase ambiguity (represented by $n_{ab}$) is included. 

Figure~\ref{fig:uncal-trans-fringes-closure-phases} illustrates geometrically the gauge-invariance of the 3-polygon closure phase for uncalibrated (all three element phases corrupted randomly, ranging from 15\arcdeg~ to 75\arcdeg) and translated (in the image plane) fringes in the left and the right panels, respectively, but the discussion applies to both scenarios equally. Both scenarios cause a displacement of the fringes and the NPCs relative to the ideal case in Figure~\ref{fig:ideal-fringes-closure-phase}. As a result, the individual principal visibility phases, $\phi_{a\lceil a+1\rfloor_3}(\lambda)$, relative to the default phase centre, $\mathcal{O}$, are differently offset relative to the ideal case. However, the closure phase, which is the sum of these three phases remains unchanged as expected. This is also clear geometrically when the phase centre is shifted to any one of the three vertices of the principal triangle (denoted by $\mathcal{O}^\prime$ in red, blue, and brown), the modified phase offset, $\phi_{a\lceil a+1\rfloor_3}^\prime(\lambda)$, corresponding to the height of the triangle from the chosen vertex to its opposite side, $\delta s_{a\lceil a+1\rfloor_3}^\prime(\lambda)$ given by Equation~(\ref{eqn:perpendicular-phase-offset-vertex}), remains unchanged compared to the ideal case. The fact that the triangle's heights are unchanged, and the orientations of its sides also remain unchanged because they only depend on the array element spacings, gives rise to the SOS conservation principle. It is important to note that the displacement of the fringes in either case is constrained to be parallel to themselves such that the triangle enclosed by the three vertices of intersection (the gray shaded region), conserves its shape, orientation, and size (SOS), and thus the area too, independent of the choice of phase centre. The only degree of freedom for the triangle that does not violate the SOS conservation principle is an overall translation. This is the image-plane geometric visualisation of the gauge-invariance of closure phase, and how closure phase is fundamentally related to the properties of the triangle enclosed by the fringe NPCs in the image plane, and not to the phase centre, calibration, or image-plane translations. And this is the geometric reaffirmation of the familiar aperture plane view that closure phase is invariant to element-based calibration.

\subsection{Relation between Areas in Aperture and Image planes}\label{sec:areas-triad}

The closure phase is seen to be intricately linked to the the geometric characteristics of the triangle determined by the fringe NPCs, encapsulated by the SOS conservation principle, which implies that the triangle's area must also be gauge-invariant. This motivates investigation of the relation between the closure phase and the areas of the triangles enclosed by the fringes and the array elements in the image and aperture planes, respectively. Indeed, it can be shown that
\begin{align}\label{eqn:CPhase-areas-triad}
    \psi_3^2(\lambda) &= 16\pi^2 A_{\mathcal{A}3}(\lambda) \, A_{\mathcal{I}3}(\lambda) \, ,
\end{align}
where, $A_{\mathcal{I}3}(\lambda)$ is the area of the triangle enclosed by the three fringe NPCs in the image plane, $A_{\mathcal{A}3}(\lambda)$ denotes the area of the triangle formed by the locations of the triad of array elements. $A_{\mathcal{A}3}(\lambda)$ is in units of wavelengths squared. The subscripts $\mathcal{I}$ and $\mathcal{A}$ in $A_{\mathcal{A}3}(\lambda)$ $A_{\mathcal{I}3}(\lambda)$ and $A_{\mathcal{A}3}(\lambda)$ denote the image  and the aperture plane, respectively, while the subscript 3 denotes a 3-polygon. $A_{\mathcal{I}3}(\lambda)$ is dimensionless as it is obtained using direction-cosine coordinates. See \ref{sec:areas-triad-v1} for a detailed derivation of this result and associated caveats. A generalisation of these closure phase relationships to a closed $N$-polygon is provided in \ref{sec:closure-phase-N-polygon}. 

Figure~\ref{fig:gauge-invariance} illustrates the quantities in this relationship. $A_{\mathcal{I}3}(\lambda)$ is denoted by the gray shaded area, while $A_{\mathcal{A}3}(\lambda)$ is the area enclosed by the array elements in Figure~\ref{fig:antpos} in wavelength squared units. In this example, $A_{\mathcal{I}3}(\lambda)\approx 1.78\times 10^{-6}$, $A_{\mathcal{A}3}(\lambda)\approx 34410.43$, and $\psi_3(\lambda)\approx -3.11$~radians, thereby validating Equation~(\ref{eqn:CPhase-areas-triad}). Thus, using coordinate geometry in the image plane, this work provides a detailed derivation of similar findings from a quantum mechanics perspective \citep{Kobayashi2010}. 

\subsection{A Geometric Reasoning for the Translation}\label{sec:translation}

Here, we provide a geometrical reasoning for the translation of the three-fringe interference pattern in the presence of one or more aperture element-based phase errors. Although fully valid in a radio interferometric context, it is easily described from an optical interferometry viewpoint (see \ref{sec:optical-interferometry}).

In the context of aperture masking in optical interferometry, the three aperture elements in Figure~\ref{fig:antpos} correspond to the small unmasked regions of a larger parabolic mirror\footnote{The mask, of course, is usually implemented in the pupil plane.}. If we assume beam combination of the type used in most aperture masking experiments, i.e., image-plane combination where pupil rescaling is the only type of pupil remapping performed, then Figure~\ref{fig:triad-fringes-v2} corresponds to the imaged fringes on the focal plane of the telescope\footnote{In the optical case, the geometric delays are set by the shape and accuracy of the parabolic surface, and sidereal tracking of the fringes is performed by moving the full telescope. In radio interferometers, the array elements in the aperture plane coherently amplify the voltages, and geometric delays and sidereal fringe tracking are performed electronically, followed by cross-correlation of voltages from different array elements [see Equation~(\ref{eqn:E-field-corr})].}.

In this picture, a distortion of the wavefront's phase at one of the unmasked apertures caused by turbulence in the propagation medium along its path, effectively translates to a simple displacement of the aperture element toward or away from the prime focus, resulting in a net path length or phase difference to the focus. We have shown that such a disturbance will shift the closed three-fringe pattern on the image plane, but will obey SOS conservation. It is easy to see why the three angles of the fringe triangle, and its orientation, are preserved, since these are predetermined by the aperture's geometry, and thus the fringes can only shift perpendicular to the fringe length, as seen earlier and described by Equations~(\ref{eqn:npc}) or (\ref{eqn:triad-npc}). While less obvious, it remains physically intuitive that the lengths of the triangle's sides are also preserved, since a phase distortion associated with a single aperture affects the visibilities on the two baseline vectors that include this aperture with equal but opposite values, so that the two fringes involved shift relative to each other in such a way that the lengths between the intersecting vertices are preserved\footnote{The arguments above are true for displacements, $\Delta D_a$, small compared to the baseline lengths.}. 

\begin{figure}
\includegraphics[width=\linewidth]{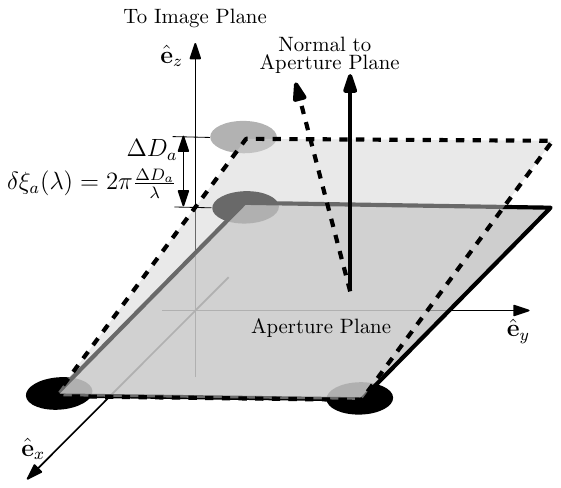}
\caption{A schematic diagram of the effect of a phase error attributable to a single element in a close triad of elements (denoted by dark circles) in an interferometer array. The original aperture plane (in dark gray shade) is at $Z=0$ with normal vector shown by the thick, solid upward arrow, with the focal (image) plane, in the $\hat{\boldsymbol{e}}_z$ direction. The phase error, $\delta\xi_a(\lambda)$, at one array element (indexed by $a$) can be effectively characterised as a change in path length, $\Delta D_a$, from that array element to the focal plane (sometimes referred to as `the piston effect' \citep{Martinache+2020}) given by $\delta\xi_a(\lambda) = 2\pi\Delta D_a/\lambda$. This change in effective path length leads to a tilt of the aperture plane (in light gray shade) as indicated by the new normal vector (tilted, dashed arrow), and hence a corresponding shift of the image plane. Thus, the image appears displaced relative to the original image plane. The fringes of all baseline vectors that contain the array element with the phase error will each be subject to a position offset as governed by Equation~(\ref{eqn:perpendicular-phase-offset}), $\Delta s_{ab}(\lambda)=\delta\xi_a(\lambda) / (2\pi|\boldsymbol{u}_{ab}|)$. Regardless of the shift, SOS conservation will apply to the three-fringe interference image. 
\label{fig:Tilt}}
\end{figure}

Figure~\ref{fig:Tilt} illustrates the result when the phase of one element in a closed triad is corrupted. The three dark circles indicate the elements (unmasked apertures in an optical telescope, or antennas in a radio interferometer) in the aperture plane (in dark shade of gray), assumed to be on the $Z=0$ plane, whose normal vector is indicated by the thick, solid upward arrow. The radiation is then directed from the elements to the focal (image) plane, wherein a three-fringe image is synthesised by the interference of the EM waves. Consider a phase corruption of one array element (indexed by $a$) by an amount $\delta\xi_a(\lambda)$, equivalent to a path length change, $\Delta D_a$, related by $\delta\xi_a(\lambda)=2\pi\Delta D_a/\lambda$, from that aperture element to the focal plane. Since three non-collinear points determine a plane, one can visualise this phase corruption, or the extra path length, at one of the aperture elements as a tilting of the aperture plane relative to the original. The tilted aperture plane and its normal are shown by the light gray-shaded region and the dashed arrow, respectively. Such a tilt then directs the light in a different direction, leading to a shift of the interference pattern in the image plane. Each of the fringes from baselines that contain the phase-corrupted aperture element will be subject to a position offset in the image plane given by Equation~(\ref{eqn:perpendicular-phase-offset}), $\Delta s_{ab}(\lambda)=\delta\xi_a(\lambda) / (2\pi\,|\boldsymbol{u}_{ab}|)$. Except for the shift, the three-fringe pattern, including the SOS characteristic, is otherwise conserved. These arguments can be generalised to a scenario even when more than one aperture element in a triad is subject to phase corruption. 

\section{APPLICATION TO REAL DATA}\label{sec:data}

We verify the visualisation and estimation of closure phase in the image plane using three examples with real data from the Karl~G.~Jansky Very Large Array \citep[VLA;][]{Perley+2009}, and from the Event Horizon Telescope\footnote{\url{https://eventhorizontelescope.org/}} (EHT). The VLA is a radio interferometer in New Mexico, comprised of 27 antennas of 25~m diameter each, arranged in a Y-pattern. The EHT is a global millimeter VLBI array involving 8 stations extending from Europe to Hawaii. 

The first example involves VLA observations of the compact radio quasar 3C~286, including both calibrated data and uncalibrated data. The second involves VLA observations of a powerful extended radio galaxy with a complex morphology, Cygnus A, using calibrated data, and then purposefully phase-corrupted data. The third example involves EHT observations at high frequencies of the active nucleus in the nearby radio galaxy, M87 \citep{Walker+2018}. 

These examples span the range from simple to complex morphologies in the image plane, and from low to high frequencies with very different phase-stability criteria for the visibilities. These examples will demonstrate the SOS conservation principle in real data under varying observing conditions and reaffirm geometrically the well known fact that the closure phase is robust to phase errors that are element-based, but not baseline-based. We will also verify that the closure phases can be estimated geometrically from the image plane, and that the results agree with those derived from the aperture plane data (visibilities) to within the estimated uncertainties. 

\subsection{Radio Quasar 3C~286}\label{sec:3C286}

The first example is that of radio quasar, 3C~286, which is a bright and highly compact object, often used for flux density and complex bandpass calibration at radio wavelengths. We employ the VLA in its largest (`A') configuration, and selected three antennas from the array, corresponding to a triangle with 
projected spacings (baselines) of 12.4~km, 7.5~km, and 15.0~km. The flux density of 3C~286 at the observing wavelength of $\lambda=3.2$~cm ($\nu=9.4$~GHz), measured on these antenna spacings is $\simeq 4.4$~Jy ($1\,\textrm{Jy} = 10^{-26} \,\textrm{W}\,\textrm{m}^{-2}\,\textrm{Hz}^{-1}$). 3C~286 is the dominant source of emission in the field of view. It has a compact core-jet structure, which on the spatial frequencies being considered herein essentially appears as an unresolved, point-like object \citep{Perley+2013}. 

The nearly point-like structure of 3C~286 implies a closure phase very close to zero, which further implies that the three fringe NPCs will intersect nearly at a point resulting in a nearly degenerate triangle. Equation~(\ref{eqn:perpendicular-phase-offset-vertex}) then implies that the height of such a triangle will be $\delta s_{ab}^\prime(\lambda)\approx 0$. 

We use a short 20~s observation made at $\lambda=3.2$~cm with a narrow bandwidth of 20~MHz. At this wavelength, the spatial frequencies (in units of number of wavelengths) are $|\boldsymbol{u}_{ab}|=(u_{ab}^2+v_{ab}^2)^{1/2}\approx 3.912\times 10^5$, $\approx 2.371\times 10^5$, and $\approx 4.749\times 10^5$, respectively. The root-mean-square (RMS) level of thermal noise in the calibrated visibilities is $\approx 33$~mJy, estimated using the VLA exposure calculator\footnote{\url{https://obs.vla.nrao.edu/ect/}} using a 2~MHz spectral channel and a 20~s averaging time interval. 

We consider both calibrated and uncalibrated data. With the former, the visibilities are expected to add coherently for a sky image, since instrumental and tropospheric phase terms at each element have been determined via a strong celestial calibrator (in this case, 3C~286 itself). The uncalibrated data includes electronics- and troposphere-induced phase offsets for each aperture element in the interferometer array, which need to be corrected via calibration before a coherent image of the target object can be synthesised.

We obtained the interference pattern from a triad of aperture elements as a dirty image (no deconvolution) made from the three visibilities using the task `\texttt{tclean}' of Common Astronomical Software Applications \citep[\texttt{CASA};][]{casa:2017} with zero iterations, which is effectively a sum of the Fourier transforms of the individual visibilities. 
There are numerous ways in which the fringe NPCs can be geometrically and directly determined from the image plane without recourse to the visibility data in the aperture plane. Here, we employed a simple method, which is neither optimal nor efficient necessarily. In the first step, we determine the intersecting vertices, $(\ell_{a\underline{b}c}, m_{a\underline{b}c})$, from the interference pattern of any pair of fringes in the image plane, typically using matched-filtering followed by peak-fitting algorithm. Of the many possible possible triangles, we preferentially choose the ones closest to the peak of the element power pattern which will yield the best signal-noise-ratio ($S/N$). Next, given this vertex and the slopes of the two fringe NPCs from the predetermined projected element spacings, $\boldsymbol{u}_{ab}$, the individual fringe NPCs that contain this intersecting vertex are determined. Finally, with the three vertices of the principal triangle determined, the closure phase can be measured geometrically using either its height [see Equation~(\ref{eqn:perpendicular-phase-offset-vertex})] or its area [see Equation~(\ref{eqn:CPhase-areas-triad})].

The thermal noise in the measurements and other systematics will lead to uncertainties in the determined fringe NPCs and thus in the measured closure phase. The phase deviations on the measured visibility phases, $\psi_{ab}(\lambda)$, from thermal noise and random systematics in a high-$S/N$ regime ($S/N\gg 1$) follow a Gaussian distribution with a standard deviation that is inversely proportional to the $S/N$ \citep{TMS2017,SIRA-II}. The corresponding position error in the fringe NPCs is given by standard error propagation between the pertinent quantities, $\psi_{ab}(\lambda)$ and $\delta s_{ab}(\lambda)$, in Equation~(\ref{eqn:perpendicular-phase-offset}) as
\begin{align}\label{eqn:phase-position-offset-error}
    \bigl[\textrm{Var}\left(\delta s_{ab}(\lambda)\right)\bigr]^{1/2} &= \frac{\bigl[\textrm{Var}\left(\psi_{ab}(\lambda)\right)\bigr]^{1/2}}{2\pi\, |\boldsymbol{u}_{ab}|} \approx \frac{(S/N)^{-1}}{2\pi\, |\boldsymbol{u}_{ab}|} \, .
\end{align}
This is also typically the case with astrometric errors in VLBI applications \citep{TMS2017}. This uncertainty will also propagate into the estimated closure phase. In the 3C~286 data analyzed here (2~MHz spectral channel, 20~s integration), the signal strength from 3C~286 and the thermal noise RMS in the visibilities are roughly uniform across the different aperture element spacings giving a $S/N\approx 133$ on each visibility. 

\begin{figure*}
\includegraphics[width=\linewidth]{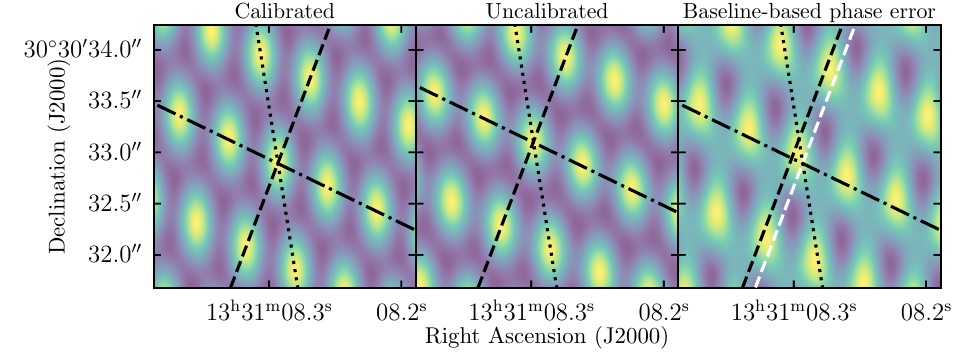}
\caption{Three-fringe interference images from calibrated (left), uncalibrated (middle), and baseline-dependent phase-corrupted (right) 3C~286 data on projected array element spacings ($\lambda\boldsymbol{u}_{ab}$) of 7.5~km (dot-dashed), 12.4~km (dashed), and 15~km (dotted) from the VLA. The image coordinates are in Right Ascension (R.A.) and Declination (Dec.) at the J2000 epoch, which are equivalent to the direction-cosine coordinates used earlier \citep{SIRA-II,TMS2017}. The principal fringe NPCs (black lines) were determined entirely from the image plane using the
method described in the text. The calibrated and uncalibrated three-fringe interference patterns look identical except that the lack of calibration shifts the interference pattern by $\approx 0\farcs2$ relative to the calibrated fringes, which indicates the magnitude of the required phase calibration terms. Independent of calibration, the principal fringe NPCs in both cases are nearly coincident with each other which geometrically confirm that 3C~286 has a highly compact structure and the closure phase, $\phi_3(\lambda)\approx 0$ as expected, remains invariant even when the element-based instrumental and tropospheric phase corruption terms remain undetermined. A baseline-dependent phase error (80\arcdeg, relative to the calibrated case) on one of the visibilities results in a shifting of the fringes corresponding to that corrupted visibility (from the uncorrupted fringe NPC shown in white dashed line to the corrupted fringe NPC in black dashed line), while the other two remain unchanged. The resulting three-fringe interference pattern (right panel) is very different from the other two panels, and the fringe NPCs are no longer coincident as evident from the non-zero area of the triangle enclosed by the three black lines, and hence, the closure phase is non-zero even for 3C~286, a point-like source. Thus, in the presence of baseline-dependent phase errors, the SOS conservation does not apply to the enclosed triangle, and the three-fringe interference image is no longer a true physical observable.  
\label{fig:3C286-all-fringes}}
\end{figure*}

Figure~\ref{fig:3C286-all-fringes} shows the three-fringe interference images made from calibrated (left panel) and uncalibrated (middle panel) data. The principal fringe NPCs are shown as black lines (dashed, dot-dashed, and dotted for 12.4~km, 7.5~km, and 15.0~km element spacings, respectively). They were determined geometrically using the simple peak-fitting procedure described above and did not use any aperture-plane measurements involving the visibilities, except the mathematically predetermined array element spacings, $\boldsymbol{u}_{ab}$ [see Equation~(\ref{eqn:triad-npc})]. For the calibrated data, the fringe NPCs nearly intersect at a point, indeed, on a grid of points, including the position of 3C~286. Importantly, the uncalibrated fringes also result in a similar grid of points. The only change is that the grid shifts by about 0\farcs2 relative to the pattern seen in the calibrated data.  

We also consider a counter-example in the right panel in which an 80\arcdeg~ phase corruption occurs in a visibility in a baseline-dependent manner rather than through one or more individual elements. In this case, two of the fringe NPCs whose visibilities were not corrupted will remain unchanged as they are unaffected by the corruption. Only the fringe NPC of the phase-corrupted baseline will be shifted (uncorrupted in white dashed line and corrupted in black dashed line). This will not result in a change of shape or orientation (which are set by the geometry of the baseline vectors), but will change the size of the triangle enclosed by the three NPCs which effectively modifies the closure phase. And the net three-fringe interference pattern appears to be very different than the calibrated and uncalibrated cases. Therefore, the closure phase (the triangle's SOS characteristics) will no longer be conserved, implying that the three-fringe interference pattern in the presence of baseline-dependent phase errors is no longer a true physical observable. This demonstrates geometrically that strict closure phase and SOS conservation only occurs if the phase error can be attributed to individual array elements (thereby affecting the visibilities in two  baselines with opposite signs), not individual baselines.

The principal closure phases were measured to be $\phi_3(\lambda)\approx 1.7\arcdeg$ and $\phi_3(\lambda)\approx 2\arcdeg$ from calibrated and uncalibrated data, respectively. The errors derived from the fitting process, based on Equation~(\ref{eqn:phase-position-offset-error}), are $\simeq 1.3\arcdeg$, implying that both results are indeed statistically consistent with zero closure phase, as expected for a point-like structure.

For verification, it is also possible to calculate the closure phase using the visibilities (in the Fourier- or aperture domain), as is typical in radio interferometry. From the individual visibility phases for each baseline in the triad, we calculate, using Equation~(\ref{eqn:CPhase-3-sum}), a closure phase of $2.6\arcdeg \pm 0.74\arcdeg$ and $2.0\arcdeg \pm 0.74\arcdeg$ for the calibrated and uncalibrated data, respectively. The RMS uncertainty in the visibility phases was again calculated as a reciprocal of the $S/N$, in radians. The closure phase is the sum of three visibility phases. Hence, the phase noise, which is uncorrelated between the three visibilities, increases by a factor of $\simeq\sqrt{3}$, to $\simeq 0.74\arcdeg$ in the closure phase. These aperture-plane estimates of closure phase are statistically consistent with that from the image plane discussed earlier. The quoted uncertainties in the aperture plane measurements represent that expected from thermal noise alone ignoring any potential systematic errors, and are thus optimistic relative to the image plane measurements.

This result demonstrates a few important principles. First, the fact that the fringes intersect at a point even for the uncalibrated data geometrically confirms the invariance of closure phase (zero, for a point-like morphology), for an instrument in which the instrumental and tropospheric phase contributions can be factored into element-based terms as in Equation~(\ref{eqn:E-field-cal-corr}). But closure phase is not invariant to baseline-dependent phase errors, which has been shown here geometrically. Second, the shift in the grid pattern in Figure~\ref{fig:3C286-all-fringes} is a measure of the magnitude of element-based phase corruptions due to the instrument and troposphere. Third, the fact that the fringes nearly intersect at a point implies that, for the VLA, the atmospheric and electronic phase corruptions to the data are predominantly factorisable into element-based gains, and are not dominated by corruptions that may be idiosyncratic to a given interferometric baseline. 
And fourth, it is possible to use this geometric understanding in the image plane for diagnosing baseline-dependent phase errors, provided a model is known for the object being imaged, or in the case of highly redundant arrays. The key distinction between element-based and baseline-based corruptions is that the former will result in an arbitrary translation of fringes in the image plane but conserve the SOS of the triangle, whereas a baseline-based error will fundamentally modify the SOS parameters of the principle triangle, depending on the nature of the baseline-dependent error. For example, a simple baseline-based phase error such as the one discussed above in the 3C286 example will modify only the size of the principal triangle, while a miscalculated antenna spacing vector can modify any or all of the SOS parameters.

\subsection{Radio Galaxy Cygnus~A (3C~405)}\label{sec:Cyg-A}

\begin{figure}
\includegraphics[width=\linewidth]{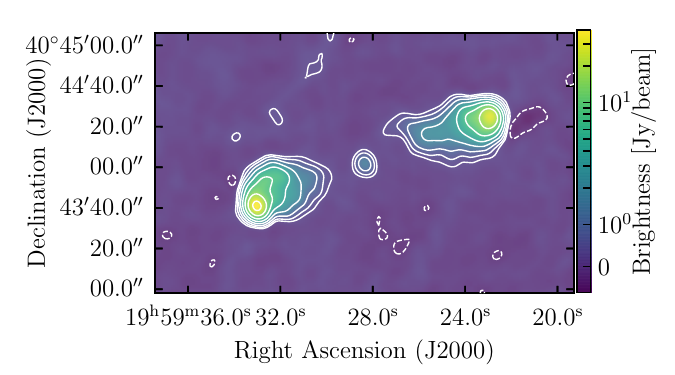}
\caption{Image of Cygnus~A, a bright radio galaxy, synthesised from 4~min and 128~MHz of VLA data at $\lambda=3.75$~cm \citep{Sebokolodi+2020}. Cygnus~A has a complex structure at these wavelengths: a bright core centred on the active galactic nucleus (AGN) and two bright and non-symmetric lobes, classified as an FR~II morphology. The angular resolution of the image (``beam size'') is $\approx 8^{\prime\prime}$. The contours correspond to $-2.5 \sigma$ (dashed), $2.5 \sigma$, $5 \sigma$, $10 \sigma$, $20 \sigma$, $40 \sigma$, $80 \sigma$, $160 \sigma$, and $320 \sigma$, where, $\sigma\approx 0.1$~Jy/beam is the RMS of noise in the image. The color bar uses a ``symmetric'' logarithmic scale to represent both negative and positive values of brightness. 
\label{fig:CygA-image}}
\end{figure}

As a second example, we employ VLA observations at $\lambda=3.75$~cm ($\nu=8.0$~GHz) of the bright, extended radio galaxy, Cygnus~A \citep{Sebokolodi+2020}. Cygnus~A has a total flux density of 170~Jy at this wavelength, distributed in two extended lobes with a full extent of $120^{\prime\prime}$. The observations were made in the `D' configuration of the VLA, which has a longest baseline of $\approx 1$~km, corresponding to a spatial resolution of $\approx 8^{\prime\prime}$. Figure~\ref{fig:CygA-image} shows an image synthesised from 4~min and 128~MHz of these data. Cygnus~A is noted to have complex spatial structure typical of an FR~II morphology (edge-brightened with bright hotspots at the outer edges of their lobes) \citep{Fanaroff+1974}. 

We choose three baselines in a rough equilateral triangle with projected baseline lengths of 797.1~m, 773.7~m, and 819.7~m, and respective correlated flux densities of 22.7~Jy, 26.4~Jy, and 38.3~Jy. We employ a single record with an integration time of 8~s and spectral channel width of 8~MHz, giving a thermal noise of $\simeq 82$~mJy in a single polarisation. 

We employ calibrated data, then corrupt the phase of one of the array elements in a closed triad by 80\arcdeg, as would occur if, for instance, there was a significant miscalibration, to detail its geometrical effect on the resulting fringes, and eventually corrupt all three element phases by random amounts ($-68.4$\arcdeg, 101.7\arcdeg, and  90.4\arcdeg) as representative of real-world conditions. From the aperture-plane visibilities, we calculate a closure phase for the calibrated and the two forms of corrupted visibilities of $112.7\arcdeg\pm 0.3\arcdeg$, where the uncertainty is set by the quadrature sum of the individual phase errors based on the respective visibility $S/N$ ($\gtrsim 275$) using Equation~(\ref{eqn:phase-position-offset-error}). 

\begin{figure*}
\includegraphics[width=\linewidth]{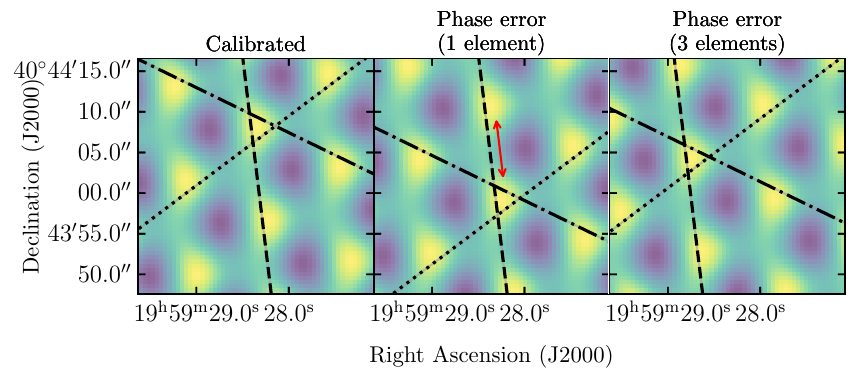}
\caption{Three element interference patterns similar to Figure~\ref{fig:3C286-all-fringes} but for Cygnus~A data. Corrupting one or all element phases results in the shifting of the interference pattern (middle and right panels) relative to the calibrated fringes (left panel), which indicates the magnitude of the required phase calibration. When a single element's phase is corrupted by 80\arcdeg (middle panel), the two fringes involving that element get affected and shift while the third fringe which is unaffected. This results in the entire triangle enclosed by the fringes sliding along the uncorrupted fringe as indicated by the red double-headed arrow while preserving its shape, orientation, and size (SOS conservation). The argument can be extended to the case when all three element phases are corrupted by random amounts as representative of real-world conditions (right panel) by applying the same logic sequentially to one corrupted element phase at a time. Independent of the degree of calibration, the principal fringe NPCs in all cases are clearly non-coincident with each other which geometrically confirms that Cygnus~A has a complex structure (see Figure~\ref{fig:CygA-image}) in contrast to 3C~286. Gray-shaded regions indicate twice the RMS uncertainties in the determined positions of the fringe NPCs as determined from Equation~(\ref{eqn:phase-position-offset-error}), but they are barely visible due to the high $S/N$ ($\gtrsim 275$) in the visibilities. The closure phase calculated from the principal triangle's heights is $\phi_3(\lambda)\approx 112.4\arcdeg$ (see \S\ref{sec:phase-center-triads}) with an RMS uncertainty of $\approx 1.5\arcdeg$, and remains invariant even after corrupting one or more element phases by large amounts. $\phi_3(\lambda)$ estimated from the area relations in \S\ref{sec:areas-triad} are $\approx 112.5\arcdeg$, $\approx 113.7\arcdeg$, and $\approx 110.9\arcdeg$ from the fringe NPCs of the three cases considered, respectively. 
These images show clearly the SOS conservation principle, that is, for a closed triad of array elements, the resulting images are a true representation of the sky brightness distribution, independent of element-based phase corruption, besides an overall translation of the pattern which does not affect the SOS conservation. If the phase error was instead baseline-dependent (not shown here), only one of the NPCs that corresponds to the affected baseline will be displaced while the other two will remain unchanged and unconstrained by this phase perturbation, thereby changing the size of the resulting triangle in the image plane, as demonstrated in Figure~\ref{fig:3C286-all-fringes} (right) for 3C~286. Thus, the SOS conservation principle will not hold for a baseline-dependent phase error. 
\label{fig:CygA-three-fringes}}
\end{figure*}

In Figure~\ref{fig:CygA-three-fringes}, we show the 3-element interference images from calibrated (left panel) and two forms of corrupted data (one element and three element corruptions in the middle and right panels, respectively). In this case, the closure phase is clearly non-zero, and hence the three fringe NPCs do not intersect in a grid of points, as for 3C~286. However, a grid pattern remains visible in the three-fringe images, and this pattern repeats exactly, with a simple shift between the three cases considered. The phase corruption of a single element in the triad as shown in the middle panel is geometrically illustrative. The two fringes involving this corrupted element are shifted, but there is no change in the third fringe. The shifting of the pattern, and thus the triangle enclosed by the fringes, will then occur parallel to the uncorrupted fringe, as indicated by the red double-headed arrow in the middle panel. When multiple element phases are corrupted (right panel), the same reasoning can be applied sequentially to all the corrupted element phases to obtain the overall shift. Between all three cases, we find the SOS conservation principle to be valid. We calculate the closure phase in the image plane using the same process as employed for 3C~286 above using the principal triangle's heights, and find it to be $112.4\arcdeg\pm 1.5\arcdeg$ on average, where the uncertainties were estimated using the uncertainties in the points of intersection determined from the peak-fitting procedure. 

The closure phases were also estimated using the relations between the areas in the aperture and image planes. For the chosen triad, $A_{\mathcal{A}3}(\lambda)\approx 1.976\times 10^8$ (in units of wavelengths squared). The corresponding image-plane areas of the principal triangles, $A_{\mathcal{I}3}(\lambda)$, are found to be $\approx 1.236\times 10^{-10}$, $\approx 1.263\times 10^{-10}$, and $\approx 1.202\times 10^{-10}$ for the three cases. Hence, the respective closure phases computed are $\approx 112.5\arcdeg$, $\approx 113.7\arcdeg$, and $\approx 110.9\arcdeg$, which are consistent with the estimates above and confirm the relations derived in \S\ref{sec:geometric-view}.

Again, although our image-plane estimate appears to have a higher uncertainty, it must be noted that our aperture-plane uncertainty calculation represents a best-case scenario assuming ideal thermal noise, ignoring imaging systematics around a bright, complex object such as Cygnus~A. The value of closure phase inferred from the image plane is not only consistent with that estimated from the corrupted visibilities in the aperture plane, but also geometrically confirms that it is indeed independent of element-based calibration. 

\subsection{Event Horizon Telescope Observations of M87}\label{sec:EHT}

As a third example, we have analyzed data provided by the VLBI-based EHT observations of the supermassive black hole in M87. This example samples a very different regime in radio interferometry, namely, much higher frequencies and much longer baselines. Therefore, the data are at a much finer spatial resolution ($\sim 20\,\mu$as), and the phase stability is more of a challenge relative to tied-array interferometry with the VLA\footnote{Tied-array implies a distributed timing signal from a central local oscillator that provides relative stability for element phases across the array. For VLBI observations spanning inter-continental baselines, such as is employed with the EHT, phase-stable local oscillator distribution is impossible, and local timing has to be maintained via accurate hydrogen maser clocks at each station in the array. The synchronisation of these clocks is one of the main sources of uncertainty in determining the interferometric phases of the array \citep[][Lecture 22 (Walker)]{SIRA-II}.}. 

The EHT data \citep{eht19-data-M87-StokesI}\footnote{\url{https://eventhorizontelescope.org/for-astronomers/data}} are described in detail in \citet{eht19-3}. In brief, observations were made of the nuclear regions of the nearby radio galaxy, M87 (Virgo~A), with the goal of imaging the event horizon of the hypothesised supermassive black hole. Observations were made on four days at 227.1~GHz and 229.1~GHz, each with a total bandwidth of 1.9~GHz, using an array comprised of seven telescopes spanning the globe, including Europe, South America, continental USA, and Hawaii. 

The publicly available EHT data have had \textit{a priori} gain (visibility flux density scale) calibration applied based on the measured system parameters at each telescope, as well as delay calibration via visibility fringe fitting, plus further adjustments based on a few redundant baselines in the array. The gain calibration provides reasonable visibility amplitudes (to within $\sim 10\%$). The delay calibration provides enough phase stability to average the data in time to 10~s records, and in frequency to a single 1.875~GHz channel. Following the EHT collaboration nomenclature, we designate these data as the `network-calibrated data'. However, the EHT collaboration emphasises that the initial calibration alone does not allow for phase-coherent imaging, since large element-based phase offsets can remain due to residual errors in the tropospheric model, station clocks, polarisation leakage, or other errors. Further element-based phase self-calibration is required to produce a phase-coherent astronomical image. They state: {\sl `Lack of absolute phase information and \textit{a priori} calibration uncertainties in EHT measurements require multiple consecutive iterations of CLEAN followed by self-calibration, a routine that solves for station gains to maximise consistency with visibilities of a specified trial image (Pearson \& Readhead 1984).'} 

\begin{figure}
\includegraphics[width=\linewidth]{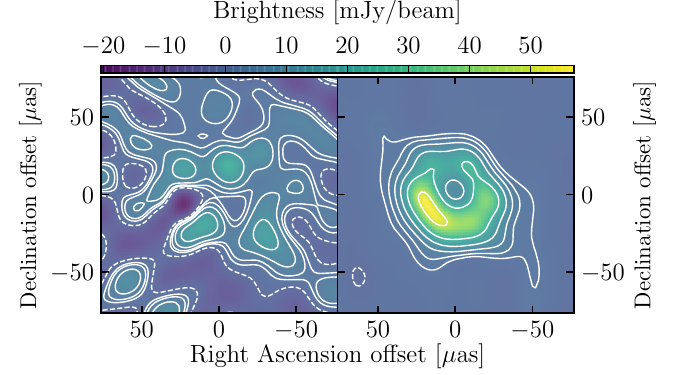}
\caption{Images of M87 at 229.1~GHz made using publicly available EHT data. \textit{Left}: Images from the `network-calibrated' data, i.e., with just \textit{a priori} flux density and delay calibration that still contains residual element-based phase errors. \textit{Right}: Images after hybrid-mapping (iterative imaging and self-calibration) as presented in \citet{Carilli+2022}. The angular resolution of the image is $\approx 20\mu$as. The contour levels of surface brightness progress geometrically in factors of two. The contours correspond to $-3 \sigma$ (dashed), $3 \sigma$, $6 \sigma$, $12 \sigma$, $24 \sigma$, $48\sigma$, and $96 \sigma$, where, $\sigma\approx 0.51$~mJy/beam is the RMS of noise in the self-calibrated image. The color bar uses a linear scale as indicated on the top. \label{fig:EHTimages_229GHz}}
\end{figure}

We have performed a standard hybrid-mapping process \citep[imaging and self-calibration;][]{Pearson+1984}, the results of which are presented in detail in \citet{Carilli+2022}. Figure~\ref{fig:EHTimages_229GHz} shows the results from our hybrid-mapping process of the EHT data \citep{Carilli+2022}. We show the images synthesised at 229.1~GHz from the network-calibrated data, and after the hybrid mapping process in the left and right panels, respectively. The former does not produce a coherent image, due to the presence of large residual element-based phase errors. After a simple hybrid mapping and self-calibration process, the image converges to an asymmetric ring with a maximum diameter of about 50~$\mu$as, consistent with the analysis of the EHT collaboration \citep{eht19-1,eht19-4}. 

Now, we present the results of the closure phase image analysis that parallels the sections above on 3C~286 and Cygnus~A. For our image-plane closure phase estimation, we select a short integration (1~min) with the most sensitive closed triad in the array, namely, the Atacama Large Millimeter Array (ALMA), the Large Millimeter Telescope (LMT), and the Submillimeter Array (SMA) stations. We then generate the three-fringe images under three scenarios: (i)~the network-calibrated data (containing residual phase errors), (ii)~the network-calibrated data, with the phase of one element further corrupted by 80\arcdeg, and (iii)~the self-calibrated data. 

\begin{figure*}
\includegraphics[width=\linewidth]{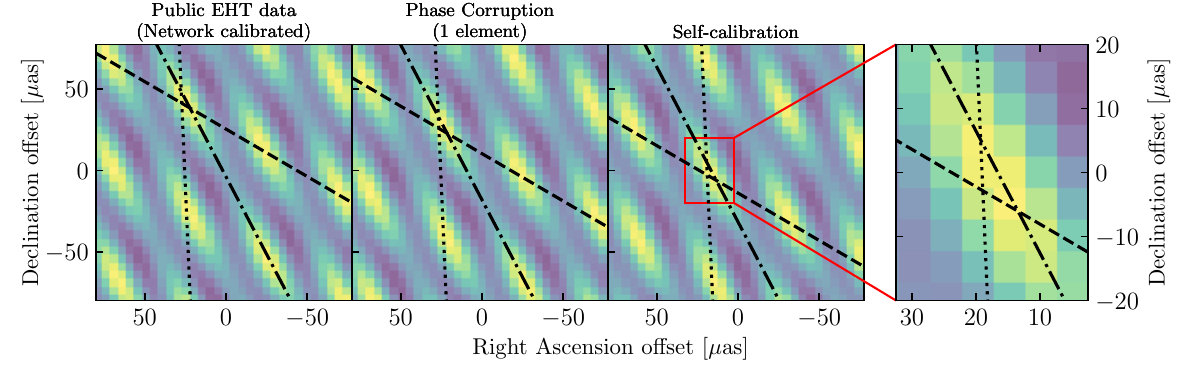}
\caption{Three-fringe interference images of M87 using a snapshot (1~min) of EHT data at 229.1~GHz. The stations involved are: ALMA, the LMT, and the SMA. The first panel (from left) shows the three-fringe interference pattern that has \textit{a priori} flux density scale and delay calibration applied. The second panel uses the same data but with one element (ALMA) phase-corrupted by 80\arcdeg. The third panel is obtained by hybrid mapping and self-calibration. The three-fringe interference pattern is found to be the same across these panels except for an overall translation relative to each other. The fourth panel is an inset showing the zoomed-in view of the self-calibrated three-fringe interference pattern. The fringe NPCs enclose a triangle of a finite area, thereby indicating a non-zero value for closure phase that was estimated from the image plane to be $\approx 38.8\arcdeg$ and $\approx 38.5\arcdeg$ from the ``principal triangle's height'' and ``product of areas'' methods, respectively. These agree with the value of $\approx 37.5\arcdeg$ derived from the aperture plane measurements (i.e., the visibilities) to within the expected uncertainties. Besides confirming that the three-fringe interference pattern remains the same except for relative overall shifts, these closure phase estimates were found to be consistent between the three panels denoting different degrees of calibration accuracy, thereby verifying the SOS conservation principle. \label{fig:EHT-3fringes}}
\end{figure*}

The results are shown in Figure~\ref{fig:EHT-3fringes}, from left to right, respectively. The three-fringe interference image in all scenarios are identical besides an overall shift, thus clearly demonstrating the SOS conservation principle. The three principal NPCs (black lines) determined geometrically in the image plane are also shown. The right panel shows a zoomed-in view of the triangle enclosed by the three fringe NPCs. From this triangle enclosing a finite area, we estimate the closure phase in the image plane by applying the methods described in \S\ref{sec:geometric-view}, similar to the 3C~286 and Cygnus~A examples above. The closure phase calculated from the principal triangle's heights are 37.9\arcdeg, 37.4\arcdeg, and 41\arcdeg~ for the three scenarios, respectively, with errors $\sim 20\arcdeg$ estimated from the fitting process. The closure phase estimates from the ``product of areas'' method are 37.8\arcdeg, 37.1\arcdeg, and 40.6\arcdeg, respectively, for the three scenarios. 

From the visibilities themselves, we derive closure phases from these three fringes to be 37.1\arcdeg, 37.1\arcdeg, and 37.9\arcdeg~ for the three scenarios, respectively. We calculate the uncertainty on these values by examining the scatter of the phases on the least sensitive baseline in the triad (LMT to SMA), over scans of 10~min to 15~min using 1~min records. The resulting phase RMS is $\sim 15\arcdeg$. This value is consistent with the $S/N$ on that visibility, which is between 4 and 5 \citep[][Lecture 9 (Wrobel \& Walker)]{SIRA-II}. We consider this to be the RMS error on the closure phase measurements, since the RMS phase scatter on visibilities involving ALMA is more than 10 times lower, and hence does not contribute appreciably to the closure phase uncertainty.

Once again, the image-plane methods are not only consistent with each other, but also with the standard aperture-plane method based on summing the visibility phases presented above. These results demonstrate that closure phases can be estimated in the image plane even in challenging interferometric experiments, such as high frequency EHT VLBI imaging of M87. Moreover, after \textit{a priori} flux density calibration, Figure~\ref{fig:EHT-3fringes} shows that the snapshot three-fringe images are good observable representations of the true sky brightness, independent of element-based self-calibration or phase corruption, besides the overall translations. This can be compared to the network-calibrated vs. self-calibrated images in Figure~\ref{fig:EHTimages_229GHz}, in which the former does not produce a coherent image. The reason is, while the snapshot three-fringe images on closed baselines may be true representations of the sky, they have independent unconstrained translations that, when summed, would not produce a coherent image. 

\section{SUMMARY}\label{sec:summary}

Closure phase has been extensively used in astronomical interferometry for decades. But, due to its inherently higher-order dependence on the moments of the spatial intensity and spatial coherence, a geometric understanding of closure phase in the image plane has been lacking. In this paper, we show how closure phase manifests itself in the image plane. We derive and demonstrate the shape-orientation-size (SOS) conservation principle in the image plane as the Fourier domain counterpart of the familiar aperture-plane closure phase. The properties of the principal image-plane triangle enclosed by the three fringes of a closed triad of array elements are preserved, even in the presence of large element-based phase errors, besides possibly an overall translation of the 3-element interference pattern. We establish that the SOS conservation principle is the precise geometric analog of the invariance of closure phase to aperture element-dependent corruptions introduced by the propagation medium and the measuring instrument, as well as any translations of the intensity distribution in the image, which has only been understood mathematically from the viewpoint of the aperture plane to date. 

The triangle SOS conservation principle provides two avenues to estimate closure phase directly in the image plane, rather than the standard practice of three measurements in the Fourier domain (or the aperture plane). First, the closure phase from a triad of aperture elements can be geometrically estimated in the image plane from a single measurement of any one of the heights of the triangle enclosed by the three principal fringes. Second, it can also be estimated from an invariant relationship that exists between the squared closure phase and the product of the areas enclosed by the triad of array elements in the aperture plane and the triangle enclosed by three fringes in the image plane. 
We have generalised these invariant relationships derived for closure phases on triads to any closed polygon in an $N$-element interferometer array. 
This geometric understanding of closure phase has potential to provide a heuristic for identifying and distinguishing between baseline-based and element-based corruptions in the image plane.

By analyzing real interferometric VLA observations of the bright radio quasar 3C~286 and the radio galaxy Cygnus~A, and high frequency EHT VLBI observations of M87, we have independently computed the closure phase using the conventional aperture-plane method and the direct geometric method in the image plane, thereby validating the latter approach. We also verify geometrically in the image plane that the closure phase in real data is robust to element-based phase errors and calibration. The results confirm the SOS conservation principle over a wide range of radio interferometric conditions. Although this geometric understanding of closure phase in the image plane (namely, the SOS conservation principle) was motivated by radio interferometry for astronomy applications, the existence of close parallels in optical interferometry as well as other disciplines has been identified \citep[appendix of][and references therein]{Thyagarajan+2020c}.

In future work, we will explore the application of SOS conservation principle to interferometric imaging where the basic output of the interferometer are images. The fact that each three-fringe image is a true representation of the sky surface brightness, independent of element-based phase calibration, presents the opportunity to perform interferometric imaging and self-calibration completely in the image plane, without conversion to aperture plane quantities. 

\begin{acknowledgement}
We acknowledge valuable inputs from Rajaram Nityananda, Arul Lakshminarayanan, David Buscher, Rick Perley, James Moran, Craig Walker, and Michael Carilli. We thank Kumar Golap for help in using the Common Astronomical Software Applications \citep[CASA;][]{casa:2017}. We thank L. Sebokolodi and R. Perley for permitting use of the Cygnus A data. We thank the EHT collaboration for making the M87 data public. We also thank the anonymous reviewer for their inputs that helped improve the manuscript. We acknowledge the use of software packages including AstroUtils\footnote{AstroUtils is publicly available for use under the MIT license at \url{https://github.com/nithyanandan/AstroUtils}} \citep{AstroUtils_software}, Precision Radio Interferometry Simulator \citep[PRISim\footnote{PRISim is publicly available for use under the MIT license at \url{https://github.com/nithyanandan/PRISim}};][]{PRISim_software}, Astropy \citep{astropy:2013,astropy:2018}, NumPy \citep{numpy:2006,numpy:2011}, SciPy \citep{scipy:2020}, Matplotlib \citep{matplotlib:2007}, Pyuvdata \citep{pyuvdata:2017}, and Python. When this work was carried out, Nithyanandan Thyagarajan was a Jansky Fellow of the National Radio Astronomy Observatory. This work makes use of the following VLA data: VLA/19A-024, VLA/14B-336. The National Radio Astronomy Observatory is a facility of the National Science Foundation operated under cooperative agreement by Associated Universities, Inc. 

\end{acknowledgement}


\appendix

\section{Derivation of Relation between Closure Phase and Areas in the Aperture and the Image Planes}\label{sec:areas-triad-v1}

In Figure~\ref{fig:ideal-fringes-closure-phase}, consider the segment bounded by the vertices of the intersection of the NPC of fringe $F_{01}(\hat{\boldsymbol{s}},\lambda)$ with the other two fringe NPCs, as the base of the triangle. Let $\theta_{0\underline{1}2}(\lambda)$ be the angle between the NPCs of the fringes $F_{01}(\hat{\boldsymbol{s}},\lambda)$ and $F_{12}(\hat{\boldsymbol{s}},\lambda)$. Then $\theta_{0\underline{1}2}(\lambda)$ is also the angle between $\boldsymbol{u}_{01}$ and $\boldsymbol{u}_{12}$ in the aperture plane. Thus, the base of the triangle is $b(\lambda) = \delta s_{12}^\prime(\lambda)/\sin\theta_{0\underline{1}2}(\lambda)$, where $\delta s_{12}^\prime(\lambda)$ is the positional offset of the NPC of the fringe $F_{12}(\hat{\boldsymbol{s}},\lambda)$ from its opposite vertex. The height is simply $h(\lambda)=\delta s_{01}^\prime(\lambda)$. Then, using Equation~(\ref{eqn:perpendicular-phase-offset-vertex}), the area enclosed by the fringe NPCs in the image plane is
\begin{align}
    A_{\mathcal{I}3}(\lambda) &= \frac{1}{2}\, \frac{\delta s_{01}^\prime(\lambda)\,\delta s_{12}^\prime(\lambda)}{\sin\theta_{0\underline{1}2}(\lambda)} = \frac{\psi_{01}^\prime(\lambda)\,\psi_{12}^\prime(\lambda)}{8\pi^2\, |\boldsymbol{u}_{01}|\, |\boldsymbol{u}_{12}|\, \sin\theta_{0\underline{1}2}(\lambda)} \, . \label{eqn:area-fringe-triad}
\end{align}
Substituting $A_{\mathcal{A}3}(\lambda) = (1/2)\,|\boldsymbol{u}_{01}|\, |\boldsymbol{u}_{12}|\, \sin\theta_{0\underline{1}2}(\lambda)$, we get 
\begin{align}
    \psi_3^2(\lambda) &= 16\pi^2 \, A_{\mathcal{A}3}(\lambda) \, A_{\mathcal{I}3}(\lambda) \, . \label{eqn:CPhase-areas-triad-v1}
\end{align}

Note that the equations in this section are directly applicable only to non-parallel fringes (or, non-collinear array elements in the aperture plane) for which $C_{a\underline{b}c}\ne 0$, or equivalently, $\theta_{0\underline{1}2}(\lambda)\ne 0$ and $A_{\mathcal{A}3}(\lambda)\ne 0$. In the limiting case when the triad of array elements are collinear in the aperture plane, $\theta_{0\underline{1}2}(\lambda)=0$, and hence, $C_{a\underline{b}c}=0$ and $A_{\mathcal{A}3}(\lambda)=0$. Because the fringe NPCs are parallel to each other and do not have a distinct point of intersection between them, the area enclosed by the fringe NPCs on the tangent-plane of the image is indeterminate, from Equation~(\ref{eqn:area-fringe-triad}). However, the product of these two areas is still a well-defined, finite value proportional to the closure phase squared, given by Equation~(\ref{eqn:CPhase-areas-triad-v1}).

\section{Closure Phase Relationships on $N$-Polygons}\label{sec:closure-phase-N-polygon}

\begin{figure}
\includegraphics[width=0.9\linewidth]{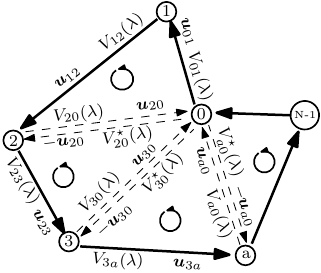}
\caption{An aperture-plane view of an $N$-polygon interferometric array, indexed by $a=0,1,\ldots,N-1$. The aperture element spacing in wavelength units (or spatial frequencies) and the corresponding spatial coherence are indicated by $\boldsymbol{u}_{a\lceil a+1\rfloor_N}$ and $V_{a\lceil a+1\rfloor_N}(\lambda)$, respectively, on the adjacent sides. By choosing a vertex (indexed by 1 in this case), adjacent triads sharing this common vertex and having one overlapping side (shown by dashed lines) with the next triad can be defined, each with its own closure phase, $\psi_{3(q)}(\lambda),\, q=1,2,\ldots N-2$. The closure phase on the $N$-polygon is the sum of the closure phases on these adjacent triads with a consistent cyclic rotation of the vertices as indicated by the arrowed circles, $\psi_N(\lambda) = \sum_{q=1}^{N-2} \psi_{3(q)}(\lambda)$. \label{fig:N-polygon-cphase}}
\end{figure}

The closure phase relations established on a triad of array elements can be extended to generic closed $N$-polygons in the aperture plane. A closed $N$-polygon can be decomposed into $N-2$ adjacent triads with each adjacent pair sharing an edge and all such triads sharing a common vertex as shown in Figure~\ref{fig:N-polygon-cphase}. The closure phase on the $N$-polygon is simply the sum of the closure phases on the adjacent elemental triads defined here, where the visibility phase measured by the element spacing on the edge shared between adjacent triads appears as the negative of each other and thus vanishes perfectly in the sum \citep[see also][]{Cornwell1987}. 

Assuming non-collinear aperture array elements, the intersection between the NPCs of fringes, $F_{01}(\hat{\boldsymbol{s}},\lambda)$ and $F_{N1}(\hat{\boldsymbol{s}},\lambda)$, can be chosen, for example, as the phase centre, $\hat{\boldsymbol{s}}_0$. Then, the visibility phases on these two fringes vanish because the NPCs of fringes $F_{01}(\hat{\boldsymbol{s}},\lambda)$ and $F_{N1}(\hat{\boldsymbol{s}},\lambda)$ pass through $\hat{\boldsymbol{s}}_0$. The closure phase on the $N$-polygon is then determined by the rest of the $N-2$ fringe NPCs. From Equation~(\ref{eqn:triad-npc}), the visibility phases of the fringe NPCs for the chosen phase centre are
\begin{align}\label{eqn:modified-intPhase}
    \psi_{ab}^\prime(\lambda) &=  
    \begin{cases}
    0 \, ,              &\, a=0,N-1 \, ,  \\
    2\pi \boldsymbol{u}_{ab}\cdot \hat{\boldsymbol{s}}_0 + \psi_{ab}(\lambda) \, , &\, \textrm{otherwise}, 
    \end{cases} \\
    \textrm{with},\, b &= \lceil a+1\rfloor_N \, , \quad\textrm{and},\nonumber \\
    \psi_{a\lceil a+1\rfloor_N}(\lambda) &= -2\pi \boldsymbol{u}_{a\lceil a+1\rfloor_N}\cdot \hat{\boldsymbol{s}} \, , \quad a=0,1,\ldots N-1 \, . \nonumber
\end{align}
Here, $\psi_{ab}^\prime(\lambda)$ is simply the phase offset proportional to the positional offset between $\hat{\boldsymbol{s}}_0$ and each of the fringe NPCs given by Equation~(\ref{eqn:perpendicular-phase-offset}). Using Equation~({\ref{eqn:triad-npc}}), the closure phase is obtained by summing the closure phases of each of the adjacent triads, which are effectively identical to the phase offsets, $\psi_{a\lceil a+1\rfloor_N}^\prime(\lambda)$, corresponding to these position offsets. Thus, similar to Equation~({\ref{eqn:modified-CPhase-3-sum}}), we get
\begin{align}\label{eqn:modified-CPhase-N-sum}
    \psi_N(\lambda) &= \sum_{q=1}^{N-2} \psi_{3(q)}(\lambda) = \sum_{a=0}^{N-1} \psi_{a\lceil a+1\rfloor_N}^\prime(\lambda) \, ,
\end{align}
where, the subscript $q$ indexes the $N-2$ adjacent triads constituting the closed $N$-polygon, and $\psi_{3(q)}(\lambda)$ denotes the closure phase on triad $q$. Equation~(\ref{eqn:modified-CPhase-N-sum}) is a generalisation of Equation~(\ref{eqn:modified-CPhase-3-sum}) for the $N$-polygon. 

Note that all the relations throughout the paper hold for any arbitrary closed polygon in any configuration between the available vertices, including self-intersecting polygons, and not limited to only the convex or concave configurations. Each polygon configuration will have a unique closure phase, in general, of course.

We now examine whether the SOS conservation property applies directly to an $N$-polygon of fringe NPCs in the image plane when $N\ge 4$. This can be understood by perturbing the phase of one of the aperture array elements. This phase perturbation will affect two fringes whose baseline vectors contain this aperture element with opposite displacements of their respective fringe NPCs. However, the rest of the fringe NPCs will remain unchanged and are unconstrained by this change. Therefore, the superposed fringe interference pattern from all the $N$ array elements ($N\ge 4$) will not be conserved on the whole. However, when the $N$-polygon is decomposed into adjacent, elemental triads, as described above, then the individual triad patterns will obey the SOS conservation principle as discussed earlier. 

The relationship established in \S\ref{sec:areas-triad} between the closure phase and the areas in the aperture  and the image planes can be extended to an $N$-polygon by expressing it in terms of adjacent and elemental $3$-polygon units as above, each of which obey gauge-invariance and SOS conservation. Let the elemental triads, all sharing a common vertex (denoted by index $a=0$) in the aperture plane, be indexed by $q=1,2,\ldots N-2$. 
As a simple example, consider a 4-polygon in the aperture plane with four vertices indexed by $a=0,\ldots N-1$, with $N=4$. The two adjacent triangles with a common vertex at $a=0$ are denoted by $\Delta_{012}$ and $\Delta_{023}$ with areas $A_{\mathcal{A}3(q)}(\lambda)$ with $q=1$ and $q=2$, respectively, in the aperture plane. The area of the 4-polygon is $A_{\mathcal{A}4}(\lambda) = \sum_{q=1}^2 A_{\mathcal{A}3(q)}(\lambda)$. Note that the edge joining the vertices 0 and 2 is only intermediate and the visibility phase on this edge will be immaterial as we will express the results using only gauge-invariant quantities from the individual elemental triads. 

\begin{figure}
\includegraphics[width=\linewidth]{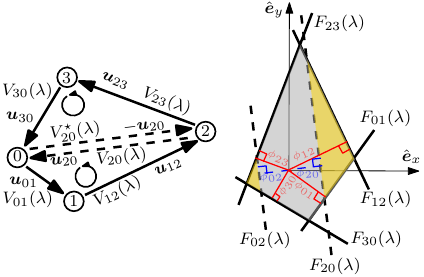}
\caption{\textit{Left}: An aperture-plane view of a 4-polygon interferometric array decomposed as two adjacent triads sharing an edge (dashed lines). The element spacing of the shared side in one triad is negative of that in the adjacent triad as indicated. Thus, the corresponding spatial coherences are conjugates of each other. The area of the 4-polygon is  $A_{\mathcal{A}4}(\lambda) = \sum_{q=1}^2 A_{\mathcal{A}3(q)}(\lambda)$. \textit{Right}: An image-plane view of the visibility phases on the 4-polygon and the adjacent triads using the principal NPCs of the corresponding fringes, $F_{ab}(\hat{\boldsymbol{s}},\lambda),\, a,b=0,1,\ldots N-1,\, b\ne a$. The principal fringe NPCs from adjacent spacings in the 4-polygon are shown by the thick, solid black lines, while that of the spacing shared by the adjacent triads is shown by the two dashed lines where one phase is negative of the other [$\phi_{02}(\lambda) = -\phi_{20}(\lambda)$] due to the conjugate relationship between their spatial coherences. The closure phases of the two triads are $\phi_{3(1)}(\lambda)=\phi_{01}(\lambda)+\phi_{12}(\lambda)+\phi_{20}(\lambda)$ and $\phi_{3(2)}(\lambda)=\phi_{02}(\lambda)+\phi_{23}(\lambda)+\phi_{30}(\lambda)$, where the visibility phases, $\phi_{ab}(\lambda)$ are the phase offsets associated with the positional offsets of the phase centre (origin) from the respective fringe NPCs according to Equation~(\ref{eqn:perpendicular-phase-offset}). The closure phase of the 4-polygon is the sum of closure phases of the two adjacent triads, $\phi_4(\lambda)=\sum_{q=1}^2 \phi_{3(q)}(\lambda) = \sum_{a=0}^3 \phi_{a\lceil a+1\rfloor_4}(\lambda)$. However, the area enclosed by the fringe NPCs of the 4-polygon (area enclosed between the four thick, solid black lines), $A_{\mathcal{I}4}(\lambda)$, is not equal to the sum of the areas enclosed by the two sets of triad fringe NPCs (the two yellow-shaded regions). Thus, $A_{\mathcal{I}4}(\lambda)\ne\sum_{q=1}^2 A_{\mathcal{I}3(q)}(\lambda)$. The SOS conservation does not apply directly to the 4-fringe pattern (denoted by their NPCs in solid black lines) as a whole. However, the SOS conservation holds individually for the elemental triad fringe patterns denoted by the yellow shaded regions. 
\label{fig:quad-fringes}}
\end{figure}

The closure phase relations apply to each of the $N-2$ adjacent elemental triads (indexed by $q$) constituting the $N$-polygon. For the 4-polygon, $q=1,2$. Thus, from Equation~(\ref{eqn:CPhase-areas-triad-v1}),
\begin{align}
    \psi_{3(q)}^2(\lambda) &= 16\pi^2 \, A_{\mathcal{A}3(q)}(\lambda) \, A_{\mathcal{I}3(q)}(\lambda) \, , \quad q=1,2 \, . 
\end{align}
Because $\psi_4(\lambda) = \sum_{q=1}^2 \psi_{3(q)}(\lambda)$ from Equation~(\ref{eqn:modified-CPhase-N-sum}),
\begin{align}\label{eqn:CPhase-areas-4-polygon-v1}
    \psi_4^2(\lambda) &= 16\pi^2\,\sum_{q=1}^2 A_{\mathcal{A}3(q)}(\lambda) \, A_{\mathcal{I}3(q)}(\lambda) \nonumber\\ 
    &\qquad\qquad + 2\,\sum_{q=1}^1\sum_{r=q+1}^2 \psi_{3(q)}(\lambda)\, \psi_{3(r)}(\lambda) \, ,
\end{align}
which can be generalised to an $N$-polygon as
\begin{align}\label{eqn:CPhase-areas-N-polygon-v1a}
    \psi_N^2(\lambda) &= 16\pi^2\,\sum_{q=1}^{N-2} A_{\mathcal{A}3(q)}(\lambda) \, A_{\mathcal{I}3(q)}(\lambda) \nonumber\\ 
    &\qquad\qquad + 2\,\sum_{q=1}^{N-3}\sum_{r=q+1}^{N-2} \psi_{3(q)}(\lambda)\, \psi_{3(r)}(\lambda) \, .
\end{align}
Alternatively, we can also express the relation between the area of the $N$-polygon in the aperture plane and the closure phases in the adjacent elemental triads as
\begin{align}\label{eqn:CPhase-areas-N-polygon-v2a}
    A_{\mathcal{A}N}(\lambda) &= \sum_{q=1}^{N-2} A_{\mathcal{A}3(q)}(\lambda) = \frac{1}{16\pi^2}\,\sum_{q=1}^{N-2} \frac{\psi_{3(q)}^2(\lambda)}{A_{\mathcal{I}3(q)}(\lambda)} \, .
\end{align}
It is noted that both Equations~(\ref{eqn:CPhase-areas-N-polygon-v1a}) and (\ref{eqn:CPhase-areas-N-polygon-v2a}) are gauge-invariant. The former expresses the closure phase on the $N$-polygon in terms of its adjacent elemental triads. The latter expresses the area of the $N$-polygon in the aperture plane as a weighted sum of closure phases on the adjacent elemental triads where the weights are inversely proportional to the areas enclosed by the fringe NPCs of the elemental triads. In either case, the gauge-invariant closure phase relations on each of the elemental triads, and hence on the $N$-polygon, can be measured geometrically. The 4-polygon example is illustrated in Figure~\ref{fig:quad-fringes}.

Note that in either of the equations above, the area under the fringe NPCs is expressed only in terms of the elemental triangle NPCs and not the $N$-polygon in the image plane. This is because the area enclosed by the fringe NPCs of the $N$-polygon is not the sum of the elemental triad fringe NPCs in the image plane, as illustrated in Figure~\ref{fig:quad-fringes}. Therefore, $A_{\mathcal{I}4}(\lambda)\ne\sum_{q=1}^2 A_{\mathcal{I}3(q)}(\lambda)$.

This inequality results from the fact that the SOS conservation is not expected to directly apply for the 4-fringe pattern. For example, perturbing the phase of array element ``1'' will only displace the NPCs of fringes $F_{01}(\lambda)$ and $F_{12}(\lambda)$ leaving the NPCs of fringes $F_{23}(\lambda)$ and $F_{30}(\lambda)$ unchanged. The resulting change in NPCs of fringes $F_{01}(\lambda)$ and $F_{12}(\lambda)$ and the lack of constraint on NPCs of fringes $F_{23}(\lambda)$ and $F_{30}(\lambda)$ will result in a distortion or shearing of the 4-fringe interference pattern (solid black lines) in the image plane shown on the right panel of Figure~\ref{fig:quad-fringes}. Hence, the SOS conservation does not apply to the 4-fringe interference pattern as a whole. However, the yellow regions denoting the 3-fringe interference patterns from the two adjacent, elemental triads will individually obey the SOS conservation property despite the non-conservation of the net 4-fringe interference pattern.

Although a detailed discussion of the propagation of measurement noise into the measured closure phases is beyond the scope of this paper and discussed in detail elsewhere \citep{Christian+2020,Blackburn+2020}, the general trends of the noise properties of closure phases on $N$-polygons can be inferred. The phase noise in the individual fringe is, in general, analytically involved but is well approximated by a Gaussian distribution in a high $S/N$ regime \citep{TMS2017,SIRA-II}. The same applies to closure phases as well \citep{Christian+2020,Blackburn+2020}. Since the closure phase of an $N$-polygon interferometric array is the sum of the $N$ individual fringe phases, 
[Equation~(\ref{eqn:modified-CPhase-N-sum})], 
the net uncertainty increases if the individual phase noises of the fringes are uncorrelated. As $N$ increases, the net uncertainty in the closure phase will tend to follow a Gaussian distribution as governed by the \textit{Central Limit Theorem}. In a high $S/N$ regime, the net uncertainty will follow closely a Gaussian distribution and grow as $\sim N^{1/2}$.

\section{PARALLELS IN OPTICAL INTERFEROMETRY}\label{sec:optical-interferometry}

We have approached this problem from the perspective of radio interferometric imaging, but the insight is applicable to optical interferometry, where measurements are made in the image plane, with particular relevance to aperture masking interferometry \citep{Buscher2015,Quirrenbach2001,mon03b}. Indeed, consideration of simple aperture masking provides further physical insight into the interpretation of closure phase in the image domain \citep{Baldwin+1986,Haniff+1989,Tuthill+2000,Cornwell1987}. 

In radio astronomy, the visibility phases are measured as the argument of the complex cross-correlation products of voltages between the elements, as per Equation~(\ref{eqn:E-field-cal-corr}), where the voltages are generated via coherent amplification of the radio signals at each element in the aperture plane. These visibility phases can then be summed in closed triangles to produce closure phases. In optical interferometry, voltages in the aperture plane cannot be captured and coherently amplified, and thus the element-pair visibilities are generated via mirrors (e.g., siderostats or unmasked regions of a larger aperture) and lenses, beam splitters, and/or beam combiners, then coherently reflect, focus, and interfere the light from different aperture elements onto a photon detector, typically a charge-coupled device (CCD), resulting in interference fringes. The phase and amplitude of the visibilities can then be extracted through a Fourier analysis of the image (using knowledge of the beam combination and reimaging optics), and closure phases are generated by summing these visibility phases \citep{Basden+2005,Pedretti+2005,Buscher2015}. 

SOS conservation for an image synthesised from a closed triad of baselines is an implicit criterion in the theory of optical speckle imaging with a non-redundant aperture mask, sometimes called triple-correlation (or triple-product or bispectrum) imaging \citep{Cornwell1987,Lohmann+1983,Weigelt+1983,Brown1978}. In a speckle imaging process, which employs exposures shorter than the atmospheric coherence time, and using a non-redundant aperture mask\footnote{A non-redundant mask ensures that only one aperture pair, or baseline, contributes to a given spatial frequency in the image plane. Without the mask, the many redundant spatial frequencies that would normally occur using the full mirror, will incoherently add in the image plane (incoherence arising from turbulent phase structure over the telescope), leading to decoherence of the measured visibility. The exception is in the high Strehl ratio regime, meaning close to diffraction-limited optics, where the element-based phase errors, or `piston phases', are small, and hence decoherence of redundant fringes is small. Such is the case for space telescopes \citep{Martinache+2020}.}, 
a Fourier transform of a given speckle image contains a set of spatial frequencies that are unique to a given aperture pair, or baseline, such that the visibilities derived can be traced directly and uniquely back to specific aperture pairs. In radio astronomy parlance, the resulting data set corresponds to an uncalibrated set of snapshot visibilities. From these, meaningful closure phases can be derived from the visibilities, and a standard hybrid-imaging and element-based self-calibration process can be performed, in which closure phase is inherently preserved \citep{Cornwell1987,Pearson+1984}.

In aperture masking in optical interferometry, and in some other applications of interferometric structure determination, the magnitude of amplitude errors in the aperture element-based complex gains, and of non-closing (i.e., baseline-based) phase errors, is negligible. In this case, the conservation of the relative positions of the NPCs on a closed triad of apertures implies a stricter conservation of the true image of the sky itself for that closed triad, except possibly an overall shift of the image.

\end{document}